\documentclass[preprint,12pt]{elsarticle}

\usepackage{amssymb}
\usepackage{amsmath}
\usepackage{textcomp, gensymb}
\usepackage{url}
\usepackage{hyperref}
\usepackage{soul}
\usepackage{tikz}
\usepackage{float}
\usepackage{lineno}

\newcommand{\beginsupplement}{%
        \setcounter{page}{1}
        \renewcommand{\thepage}{S\arabic{page}}
        \setcounter{table}{0}
        \renewcommand{\thetable}{S\arabic{table}}%
        \setcounter{figure}{0}
        \renewcommand{\figurename}{SFigure}%
     }

\journal{JMPS}

\begin{document}

\begin{frontmatter}



\title{Compressive Instabilities Enable Cell-Induced Extreme Densification Patterns in the Fibrous Extracellular Matrix: Discrete Model Predictions}

\author[1]{Chrysovalantou Kalaitzidou\corref{cor1}}

\author[2,3]{Georgios Grekas}
\author[1]{Andreas Zilian}
\author[3,4,5]{Charalambos Makridakis}
\author[3,4]{Phoebus Rosakis \corref{cor1}}

\cortext[cor1]{Corresponding Authors: P.R. rosakis@uoc.gr, C.K. chrysovalantou.kalaitzidou@uni.lu}

\address[1]{Department of Engineering, Faculty of Science, Technology and Medicine, University of Luxembourg, Luxembourg}
\address[2]{Aerospace Engineering and Mechanics, University of Minnesota, Minneapolis, MN, USA}
\address[3]{Institute of Applied and Computational Mathematics, Foundation for Research and Technology-Hellas, Heraklion, Greece}
\address[4]{Department of Mathematics and Applied Mathematics, University of Crete, Heraklion, Greece}
\address[5]{Department of Mathematics, MPS, University of Sussex, Brighton, UK}

\begin{abstract}
Through modelling and simulations we show that material instabilities play a dominant role in the mechanical behavior of the fibrous collagen Extracellular Matrix (ECM), as observed when the ECM is deformed by contractile biological cells. We compare two families of fiber network models, Family 1 with stable and Family 2 with unstable force-stretch response of individual fibers in compression. The latter is a characteristic of post-buckling of beams with hierarchical structure. Our simulations reveal different compression instabilities at play in each family, namely, fiber collapse (buckling) in Family 2  and fiber element collapse (snap-through) in Family 1. These result in highly localized densification zones consisting of strongly aligned fibers emanating from individual contractile cells or joining neighbouring cells, as observed  in experiments. Despite substantial differences in the response of the two families, our work underscores the importance of buckling/compression instabilities in the behavior of fibrous biological tissues, with implications in cancer invasion and metastasis.
\end{abstract}

\end{frontmatter}


\section{Introduction}
\label{S:Intro}
Cellular processes constitute a fundamental system of complex cascades of intracellular signalling pathways and biomechanical interactions between cells and their environment. Cells continually remodel the extracellular matrix (ECM) through chemical and mechanical signals \cite{Lu2011, Winkler2020}. The interplay between cells and the ECM is of great importance as it regulates a number of cellular processes \cite{Discher2005, Peyton2007}, such as cell motility and migration in physiological \cite{Lo2000} and pathological \cite{Wells2005,Provenzano2006, Provenzano2008, Riching2014} conditions, stem cell differentiation \cite{engler2007, Gattazzo2014}, as well as cell and tissue morphology \cite{Harris1981,Stopak1982, Yeung2005}.
The intrinsic actomyosin machinery enables cells to contract, thereby exerting tractions to the fibrous ECM, which results in the generation of spatial patterns of localized deformation \cite{Stopak1982, Vader2009, Shi2014, Notbohm2015, Grekas2021}. Although the mechanism for their formation has not been clarified yet, there is sufficient evidence for their implication in intercellular mechanical communication. These patterns are characterized by fiber alignment and severe material densification, localized within \textit{tethers} that join neighbouring cell assemblies \cite{Harris1981,Shi2014}, such as tumors. The tethers are densified bands where matrix density can be three to five times higher compared to the rest of the matrix. Cells were spotted to leave their cluster and advance along the axis of the tether towards the neighbouring cluster \cite{Harris1981,Shi2014}. In addition, individual cells in collagen induced fiber alignment and network densification along lines connecting them \cite{Vader2009}, while experiments with isolated fibroblasts reported that cells grew protrusions along the generated tether and towards each other \cite{Notbohm2015}. Moreover, cell-induced ECM remodelling underlies additional facets in tissue biology. Cancer studies have demonstrated the preference of tumor cells to invade along densified regions of ECM \cite{Provenzano2008, Yu2011, Ferruzzi2019}, while aligned collagen matrix serves as a highway-path which they use in order to migrate \cite{Provenzano2006, Provenzano2008}. 

In essence, every tissue component---cells and fibers---is a biomaterial with unique mechanical properties that responds accordingly to physical cues. The mechanical behavior of the ECM is thus attributed to that of its individual fibers. Previous studies \cite{Janmey2007, Wen2013} revealed the nonlinear elastic behavior of fibers, which explains why cell-induced deformations extend substantially far and are not confined to the cell boundary \cite{Notbohm2015, Rudnicki2013}. This nonlinearity is manifested by strain stiffening in tension \cite{Wen2013, Storm2005,vanderRijt2006,hudson2010, Piechocka2010} and buckling in compression \cite{munster2013, Kim2014, Notbohm2015,burkel2017}. These effects have been extensively investigated before  \cite{Notbohm2015,Rosakis2015, Abhilash2014,Ronceray2015,Xu2015,Liang2016,Grimmer2018,Sopher2018,Mann2019, Goren2020}, wherein fibers are modelled as homogeneous beams with stable (monotonically increasing) force-stretch responses. This would seem to suggest that the whole matrix (fiber network) would exhibit stable mechanical behavior. It might come as a surprise that this is actually not true even in the case of linear elastic fibers: in \cite{Friesecke2002} it is shown that the energy of such a network is a nonconvex, multi-well function of nodal displacements, which implies instability. In our recent work \cite{Grekas2021} instability is also encountered in a continuum model, obtained from orientational averaging of individual fiber response, even when the latter is stable. This model predicts localized densification patterns that bear a strong resemblance with experiments \cite{Stopak1982, Shi2014}.

Rather than homogeneous rods, ECM fibers have a bundlelike morphology characterized by a complex hierarchical structure \cite{Piechocka2010, Burla}. This assembly gives rise to unexpected mechanical effects. Subjected to large strains, fibers can be extremely extensible without breaking \cite{Burla} and they stiffen with increasing tension \cite{vanderRijt2006, hudson2010}. Especially interesting is the case when fibers are subjected to compressive forces under which they buckle, losing stiffness and eventually collapsing. Experiments with elastic fiber networks \cite{Lakes1993} and later on with fibrin \cite{Kim2015} revealed multiple regimes of stress-strain response marked by a non-linear softening instability in compression coupled with network densification. Additionally, uniaxial compression experiments on hierarchical beams \cite{Tarantino2019} revealed a  transition from hardening (positive slope in stress-strain response) to softening (negative slope) with increasing compression, while they were reversible upon load release. These studies highlight that fiber mechanics is by far more complicated, exhibiting unstable behavior that deviates from what has been addressed so far towards understanding ECM-related deformations.

Considering the above, our objective here is to develop a discrete network model that accounts for the unique intrinsic features of fiber morphology and mechanics and investigate the role of compression instability in deformation patterns associated with ECM densification, tether formation and fiber alignment. We implement two different families of fiber constitutive relations, with distinct nonlinearity and stability features. Family 1 displays a positive but decreasing stiffness with increasing compression, while Family 2 entails a stretch instability phase, where stiffness becomes negative at extreme compressions. Family 1 represents the traditional view of post-buckling behavior. Family 2 is a more radical model, incorporating recent experimental observations on buckling of hierarchical beams \cite{Tarantino2019}. Parts of this work have appeared in \cite{Kalaitzidou2019}.

We perform extensive simulations of a fiber network containing one or more contracting circular cells, in the spirit of \cite{Notbohm2015}. Simulations with our Family-2 models predict ECM densification, the formation of experimentally observed densified tethers between pairs of contracting cells and enhanced fiber alignment localized within the tethers. 
We show that ECM densification and fiber alignment are simultaneous consequences of fiber compression instability inherent in Family 2. These phenomena are stronger and highly localized within more sharply defined zones compared to predictions of previous models with stable stretch response \cite{Notbohm2015, Abhilash2014,Ronceray2015,Xu2015,Liang2016,Grimmer2018,Sopher2018,Mann2019, Goren2020}. Surprisingly, densified zones with aligned fibers also appear in Family-1 simulations, albeit with important differences. First, they require much higher levels of cell contraction and second, they reveal a different instability mechanism, namely snap-through buckling of larger fiber groups, such as triangular elements. 
 This distinct compression instability mechanism is also evident in simulations with the linear model and has not been addressed by previous studies involving discrete networks. These observations show that the presence of compression instability is critical and essential for localized densification and fiber alignment. 
 
\section{Methods}
\label{S:Methods}
We have developed a 2D discrete model of a fiber network representing the natural ECM to explore its mechanical behavior under cell induced loading. In particular, we partition a circular domain into triangular elements. Each of the three sides of an element represents an individual fiber (Fig.1a). Triangle vertices are the nodes of the network, where fibers terminate, so that \textit{fiber length} corresponds to the length of the segment between two nodes. The nodes are modeled as movable frictionless hinges with two degrees of freedom. In general, nodal displacements change the length of fibers, which are modelled as nonlinear elastic springs. As a result, the total energy of the network (sum of individual spring energies) is a function of all nodal displacements. This leads to the problem of minimizing the total energy over nodal displacements. This problem is solved numerically using advanced optimization techniques; see Subsection 2.3 below. 

\subsection{Mechanical properties}
For a single fiber, we introduce the \textit{effective stretch} $\lambda$, which equals the distance between its endpoints divided by its undeformed, or relaxed, length. The energy of a single fiber can be written as $W(\lambda)$ as a function of effective stretch $\lambda$. When the fiber is in tension, it is straight and $\lambda$ equals the actual stretch (strain +1), while $W(\lambda)$ equals the elastic energy due to stretching of the fiber. When the fiber is in compression, it may be buckled, in which case the elastic (mostly bending) energy of the fiber can still be expressed as a function $W(\lambda)$ of the distance between its endpoints, hence of the effective stretch $\lambda$. See Fig.1b. In that case $W(\lambda)$ is chosen to embody the post-buckling response of the fiber. If the deformed-position vectors of the fiber end points (nodes) are $\textbf{x}_{i}$ and $\textbf{x}_{j}$ and the undeformed fiber length is $l_{ij}$, the energy of the fiber is
\begin{equation}
\label{Wlamda}
    W \left( \frac{|\textbf{x}_{i} - \textbf{x}_{j}|}{l_{ij}} \right),
\end{equation}
the quantity within parentheses above being the effective stretch $\lambda$ of the fiber between nodes \textit{i} and \textit{j}.

We explore various models of the mechanical behavior of fibers, characterized by the force-stretch relation of a single fiber $S = S(\lambda)$, where
\begin{equation}
\label{Slamda}
    S(\lambda) = dW(\lambda)/d\lambda,
\end{equation}
is the fiber force, nondimensionalized after dividing by a coefficient with dimensions of force. These models were designed to capture fiber stiffening in tension $(\lambda > 1)$ \cite{vanderRijt2006, hudson2010}, but also softening in compression $(0 < \lambda < 1)$ due to buckling \cite{Notbohm2015, Rosakis2015, Grekas2021}. For direct observation of buckled fibers in collagen networks see \cite{munster2013, burkel2017}. We introduce two families of models. Family 1, shown in Fig.1c,

\begin{equation}
\label{Fam1}
    \textnormal{Family 1:} \quad S = S_{1k}(\lambda) = \lambda^{k} - 1, \quad k = 1,3,5,7
\end{equation} which includes the linear case (k=1) and Family 2 shown in Fig.1d:
\begin{equation}
\label{Fam2}
    \textnormal{Family 2:} \quad S = S_{2k}(\lambda) = \lambda^{k} - \lambda^{k-2}, \quad k=5,7.
\end{equation}
While for all models there is stiffening in tension (except for the linear one $S_{11}$), the difference between the two constitutive families is in compression. In all nonlinear models in Family 1, force $S$ and stiffness $dS/d\lambda$ both increase monotonically with increasing stretch $\lambda$ (Fig.1c). However, stiffness decreases monotonically with compression (as $\lambda$ decreases to zero) until it vanishes in the crushing limit $\lambda \to 0$, whereas force reaches a plateau where it remains approximately constant as $\lambda \to 0$. Models differ in how abrupt the loss of stiffness is and at which level of compressive stretch it occurs. Thus Family-1 fibers can sustain a limited amount of compressive force even after buckling. See also \cite{Grekas2021} where similar fiber behavior is used to derive a continuum model. In contrast \cite{Notbohm2015} use a bilinear model with piecewise constant stiffness that is lower in compression. Loss of stiffness in Family-1 type force response is intended to model the post buckling behavior (load versus distance between ends) of nonlinear elastic bars in compression by axial loads in the context of large deformations. Family-1 behaviour is consistent with experiments and simulations of post-buckling in certain homogeneous nonlinear elastic beams \cite{coulais2015}. 

Actual ECM fibers are far from homogeneous, but exhibit a bundle-like morphology with a complex hierarchical structure \cite{Piechocka2010,Burla}. This gives rise to unexpected mechanical effects. In a recent study \cite{Tarantino2019} with hierarchical beams, uniaxial compression experiments revealed a post-buckling response where the force-stretch relation changes from positive to negative stiffness (slope) for high enough compression. This also occurs in beams composed of metamaterials \cite{coulais2015}. Family-2 models were designed to capture this instability (Fig.1d). Stiffness decreases monotonically with decreasing $\lambda < 1$ as the fiber initially resists the compressive load, after which stiffness becomes negative with further compression, entering a compression instability regime (negative stiffness) up to final collapse as $\lambda \to 0$.

Suppose one of the models from (\ref{Fam1}) or (\ref{Fam2}) has been chosen. The corresponding elastic energy of a fiber is then given by
\begin{equation}
\label{WDef}
    W(\lambda) = \int_{1}^{\lambda} S(\gamma) \,d\gamma
\end{equation}
with a minimum at $\lambda = 1$ when the fiber is unstretched. Let $\lambda_{j}$ be the stretch of fiber j, where \textit{j} = 1,2,..., \textit{F} and \textit{F} is the total number of fibers in the network. Therefore, the total fiber network strain energy is equal to
\begin{equation}
\label{StrainEne}
    E(\textbf{x}_{1},..., \textbf{x}_{\textit{N}}) = \sum_{k=1}^{F} W(\lambda_{k}) = \sum_{i=1}^{N}\sum_{j=1}^{N} k_{ij}W \left( \frac{|\textbf{x}_{i} - \textbf{x}_{j}|}{l_{ij}} \right).
\end{equation}
Here \textit{F} is the number of fibers, \textit{N} the number of nodes and $k_{ij} = 1$ if there is a fiber joining nodes \textit{i} and \textit{j}, 0 otherwise.

\subsection{Nonconvexity and Instability}
Our initial attempt is to impose displacements or applied forces on boundary nodes of the network and minimize the total network energy $E(\textbf{x}_{1}, \ldots , \textbf{x}_{N})$ with respect to all positions of interior nodes $\textbf{x}_{i}$. However, allowing for large contractile deformations can result in nonphysical solutions due to interpenetration of matter. This happens when triangular elements fold over and snap through to the other side, as there is no resistance against fibers crossing through each other in the model. Examples are shown in Supplementary SFig.1I. This is an instance of the well-known snapthrough instability of structural mechanics, e.g., \cite{pecknold1985} and shows that the energy $E$ is nonconvex and likely to have multiple local minima, as well as unstable extrema, even in the case of the linear fiber model $S_{11}$($\lambda$)$ = \lambda -1$ as schematically shown in Fig.2a,b. This and other instabilities also occur in a regular (lattice) network of linear springs \cite{Friesecke2002}.

Solutions with snapthrough involve interpenetration of matter and orientation reversal, which are both physically unacceptable. In order to exclude such solutions, an energy penalty term is introduced that resists any two fibers with a node in common from crushing into each other. That would be equivalent to the oriented area of the associated triangular element going to zero, then becoming negative with orientation reversal. Interpolating the deformation from nodes to the entire domain in a piecewise affine way (continuous overall and linear in each triangle), we define $J$ to be the Jacobian determinant of the deformation. Hence $J$ is piecewise constant and equal to the \textit{ratio of deformed to undeformed oriented triangle area}. Letting \textbf{x} and \textbf{z} be two undeformed vector sides of a triangle, and $\bar{\textbf{x}}$, $\bar{\textbf{z}}$ be the deformed sides, we have

\begin{equation}
\label{Jacobian}
    J = \frac{(\bar{\textbf{x}} \times \bar{\textbf{z}}) \cdot \textbf{k}}{(\textbf{x} \times \textbf{z}) \cdot \textbf{k}}
\end{equation}
in terms of the vector product, where \textbf{k} is the out-of-plane vector. Negative $J$ denotes orientation reversal of the respective elements, i.e. folding over and interpenetration of matter. The penalty term is chosen as a function of $J$,
\begin{equation}
    \label{Phi}
    \Phi(J) = e^{-Q (J - b)}
\end{equation}
where $b>0$ is a small constant and $Q>0$ is large (Supplementary SFig.1II, we usually choose Q=50 and b=1/4). Thus $\Phi(J)$ is very small for $J>b$ and becomes very large for $J<0$. Thus it serves in maintaining positive orientation in the network, as negative orientation $(J<0)$ is costly in energy. For elements with positive area ratio, $\Phi(J)$ is very small, thereby having essentially no contribution to the network’s total energy. Physically it corresponds to fibers (which actually have nonzero thickness) resisting being crushed together when network elements are close to collapsing. The modified network potential energy has the form:
\begin{equation}
\label{TotalStrainEne}
    \hat{E} = \sum_{j=1}^{F} W(\lambda_{j}) + \sum_{k=1}^{K} A_{k} \cdot  \Phi(J_{k})
\end{equation}
where $F$ is the total number of fibers in the network, $K$ the total number of elements, $W(\lambda_{j})$ the potential energy of an individual fiber, $A_{k}$ the reference area that element $k$ occupies and $J_{k}$ its oriented area ratio. Even after modifying the energy by adding the penalty term, instabilities due to nonconvexity are still present. We identify some instability modes next:
\newline
\newline
\textit{Nonconvexity due to Large Rotations}
\newline
Even if the single fiber energy $W(\lambda)$ from (\ref{WDef}) is strictly convex with a minimum at $\lambda = 1$, the corresponding energy in (\ref{Wlamda}) is a nonconvex function of nodal positions $\textbf{x}_{i}$ because of rotational invariance (Fig.4 in \cite{Friesecke2002}). This is a source of nonconvexity of the total energy $\hat{E}$, that is typically entirely missed when small rotations are assumed \cite{Friesecke2002}.
\newline
\newline
\textit{Element Collapse Instability}
\newline
Before an element undergoes snapthrough (its oriented area changes signs) it buckles, or collapses, when a node touches the opposite side. This is actually an unstable equilibrium of the triangle energy which is thus a nonconvex function of element oriented area ratio $J$. Compressing an element triangle along its height keeping its base fixed (Fig.2a), we observe that the energy of the triangle as a function $W(J)$ is minimal and vanishes at $J = 1$ and at $J = -1$ (Fig.2b) after the triangle has snapped through to its mirror image. Since $W(J)$ is odd it must be nonconvex with an unstable equilibrium at $J = 0$. In our model, such total collapse is prevented by the penalty term $\Phi(J)$, but bistability and noncovexity of the penalized energy $W(J) + \Phi(J)$ remains (Fig.2c), with an additional, highly compressed solution for some values of compressive force (Fig.2d). This occurs in both model Families, even in the linear model; it is an example of the well-known snapthrough instability of structural mechanics, e.g., \cite{pecknold1985}. In order to identify this instability mode in our simulations we define the \textit{densification ratio} $\varrho$ of each triangular element to be the ratio of deformed to reference density of a hypothetical continuum deforming as the triangle. This gives
\begin{equation}
\label{densratio}
    \varrho = 1/J.
\end{equation}
\newline
\newline
\textit{Fiber Collapse Instability}
\newline
In Family-2 networks there is an additional instability: when a fiber is compressed past the point where the slope of the $S(\lambda)$ curve becomes negative (Fig.1d), it enters an unstable regime, tending to collapse to zero effective stretch. For example, the unstable regime for  $S$($\lambda$) $= \lambda^{7}$ $- \lambda^{5}$ is $0 < \lambda \leq 0.85$.
Clearly, fiber collapse would imply area collapse of any element (triangle) with this fiber as a side (Fig.2e). Eventually the penalty term (\ref{Phi}),(\ref{TotalStrainEne}) restabilizes the element against total collapse. An unstable regime remains in general, rendering fiber compression response essentially biphasic, similar to Fig.2d.

To summarize, all fiber networks are susceptible to element collapse (triangle buckling) instability. Family-2 networks suffer from an additional fiber collapse instability brought about by total loss of strength due to buckling of hierarchical fibers. Simple geometry shows that fiber collapse implies element collapse (Fig.2e), but not vice versa (Fig.2f).

\subsection{Formulation, Software and statistical analysis}
By expressing stretches $\lambda_{j}$ and Jacobians $J_{k}$ in terms of variable nodal positions $\textbf{x}_{i}$, $i = 1,... , N$, we express the total energy $\hat{E}$ in (\ref{TotalStrainEne}) as a function $\hat{E}(\textbf{x}_{1},..., \textbf{x}_{N})$ of nodal positions. See (\ref{StrainEne}) for the first term. We then perform energy minimization on $\hat{E}$.

The discrete model has been implemented in Python \cite{python}. The triangulation has been implemented in FEniCS \cite{fenics} and the optimization method (for finding energy minima) is explained in \cite{Grekas2021}. The statistical analysis has been done in R \cite{R} and for multi-group comparisons we used one-way analysis of variance (ANOVA).

\subsection{Model Geometry} 
Cells are modelled as circular cavities of radius $r_{c}$ within the domain of outside radius $R$. Thus, the domain containing the ECM network is an annulus with $r_{c} < r < R$, where  $r = |\textbf{x}|$ is the radial distance from the domain center and $\textbf{x}$ is the position vector. In particular, we model contractile cells. We simulate contraction by prescribing an inward radial displacement of cell boundary nodes given by:
\begin{equation}
\label{eq:ur}
\textbf{u}(\textbf{x}) = - \frac{u_{0}}{r_{c}}\textbf{x}
\end{equation}
for nodes with position vector $\textbf{x}$ such that $|\textbf{x}| = r_{c}$.
Simulations with two cells involve two distinct cavities with contractile displacement applied on the boundary of each. The outer boundary $|\textbf{x}| = R$ of the network is free (no applied forces or prescribed displacements) in all simulations.

\begin{figure}[H]
    \includegraphics[width=13cm]{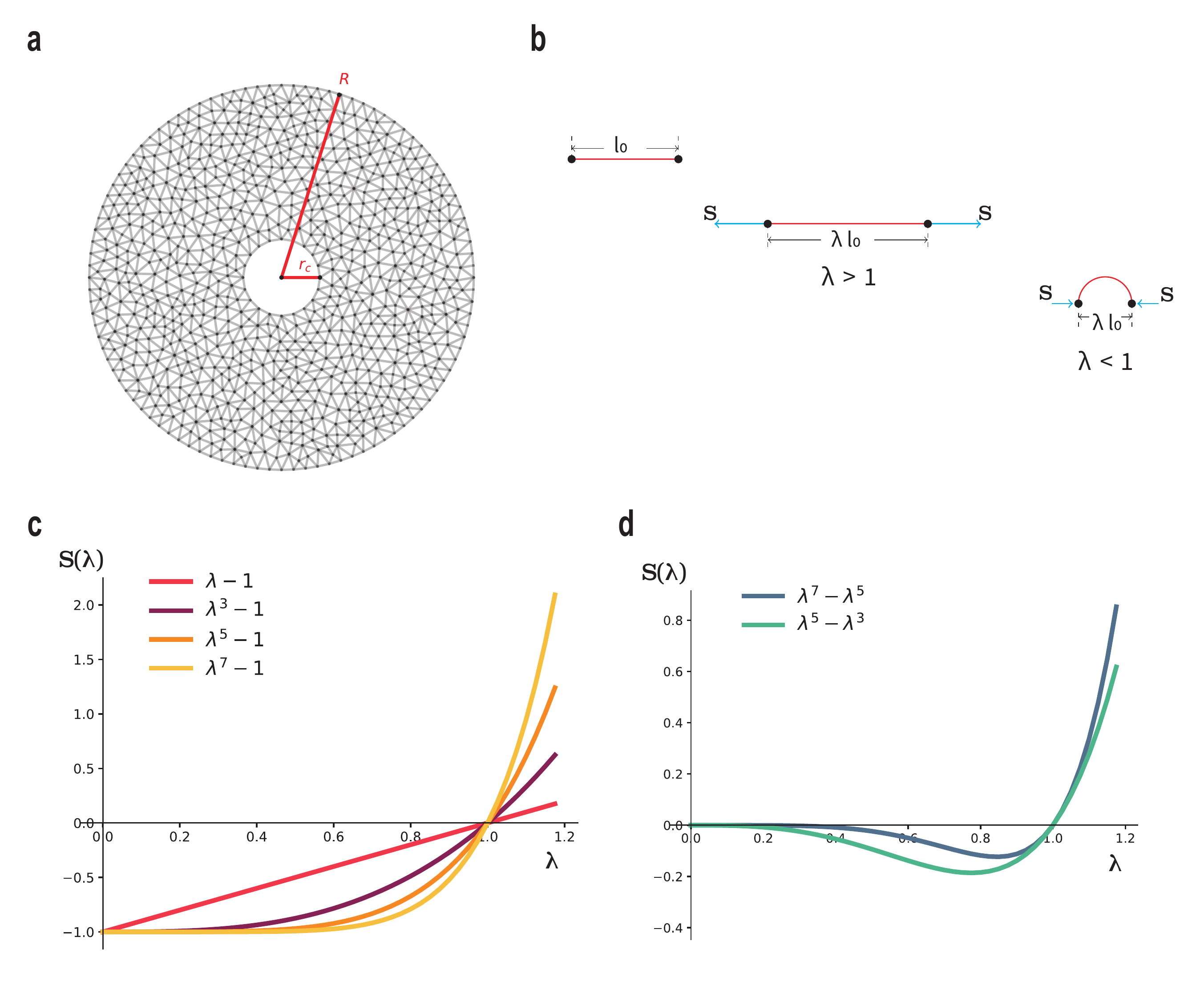}
    \label{fig:methods1}
    \caption{\textbf{(a)} Example of our 2D discrete fiber networks. Each edge corresponds to an individual fiber. The cavity represents a cell with undeformed radius $r_{c}$. \textbf{(b)} Effective stretch $\lambda$ of a single fiber. Here $\lambda$ is defined as the ratio of deformed to reference ($l_{0}$) distance of a fiber's endpoints. From left to right: a relaxed fiber with length $l_{0}$, a fiber under tension ($\lambda > 1$) and a buckled fiber under compression ($\lambda < 1$). The cyan arrows represent the applied loads at the fiber's endpoints. \textbf{(c,d)} Force-stretch ($\lambda$) curves of individual fibers. \textbf{(c)} Family 1: $S_{1k}(\lambda) = \lambda^{k} - 1$, $k = 1,3,5,7$ \textbf{(d)} Family 2: $S_{2k}(\lambda) = \lambda^{k} - \lambda^{k-2}$, $k = 5,7$.}
\end{figure}

\begin{figure}[H]
    \includegraphics[width=13cm]{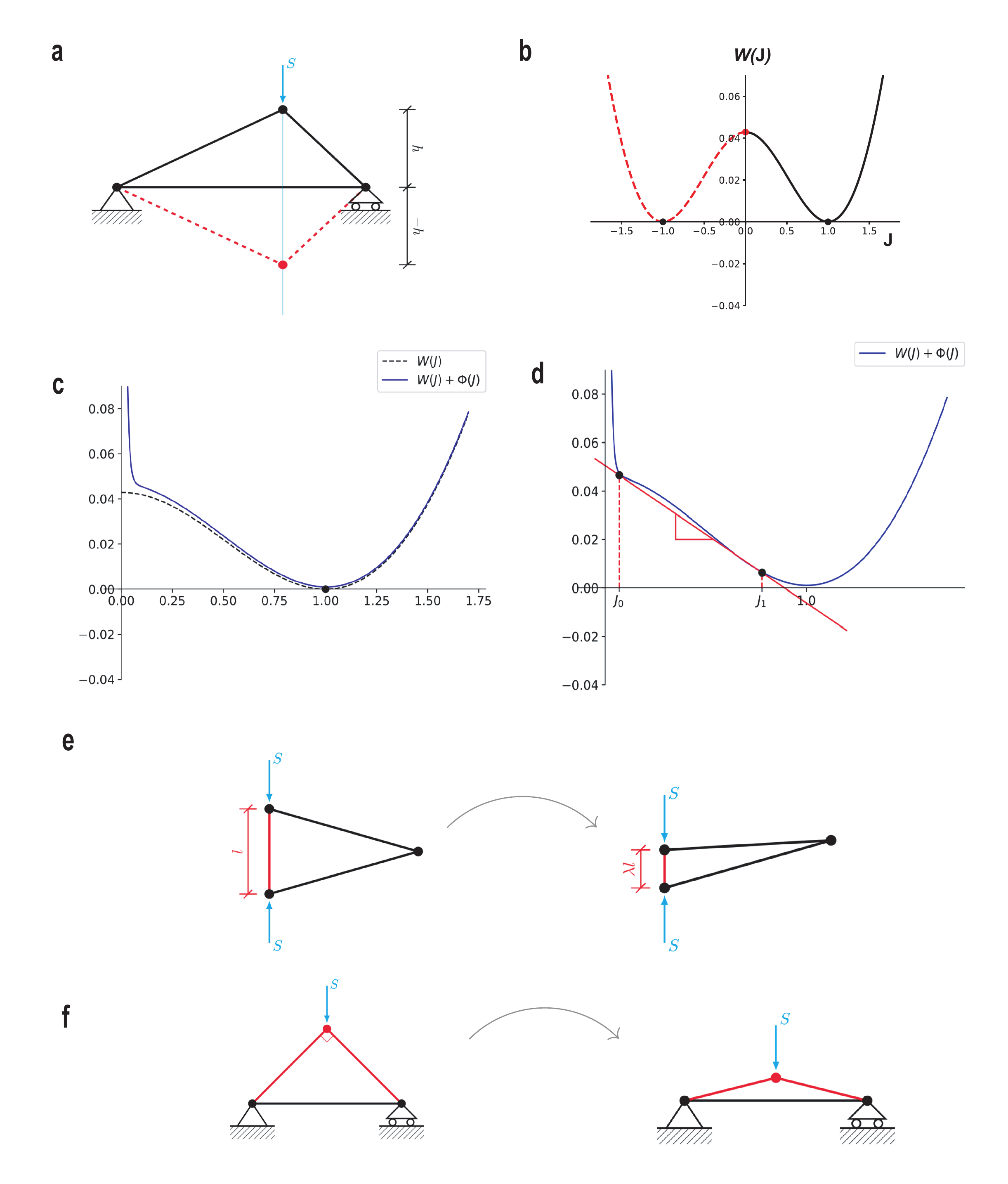}
    \label{fig:methods2}
    \caption{\textbf{ Instability Mechanisms. (a)} Element collapse Instability (snapthrough of triangular elements) under compressive force $S$ (cyan) \textbf{(b)} Energy of a triangular element as a function of its oriented area ratio $J$. Note nonconvexity and two-well structure. \textbf{(c)} Dotted line: as in (b) for $J>0$. Solid line: energy with penalty $\Phi(J)$ added. \textbf{(d)} Penalized energy has two stable equilibria $J_0$ and $J_1$  under suitable compressive force (equal to the slope of the red straight line) \textbf{(e)} Fiber collapse (red fiber) causes triangular element area collapse. \textbf{(f)} The converse is not true. Triangular element collapse can happen without fiber collapse.}
\end{figure}

\section{Results}
\subsection{Single-cell simulations}
\subsubsection{Severe localized densification patterns are observed in Family-2 models, moderate ones for Family 1}
\label{results:single1}

We simulate a single cell contracting within a fibrous network for each one of the models introduced in \eqref{Fam1}, \eqref{Fam2}. Simulations  at $50\%$ cell contraction exhibit patterns of highly localized, severe densification shown in Fig.3a-c, SFig.2a-c. These patterns take the form of bands, emanating from the periphery of the contracting cell into the surrounding matrix.  Plotting the densification ratio of each element versus distance from the cell  shows that in Family-2 models, highly densified elements have  densification ratio $\varrho \approx 3$ and  reach up to six cell radii into the ECM (Fig.3c).  In contrast,  Family-1  densified triangles are confined within  two cell radii (Fig.3a,b), with densification ratio $\varrho$  at most $2$.

The distribution of fiber stretches within the deformed networks illustrates similarities and differences between models (Fig.3d-k, SFig.2). Fibers under tension ($\lambda > 1$) align roughly with the radial direction, forming continual paths that propagate a few cell diameters into the matrix (Fig.3d-f, SFig.2d-f). This happens regardless of the model, though in Family-2 simulations the paths extend further into the matrix (Fig.3f, SFig.2f). When it comes to compressed fibers, things differ significantly between models (Fig.3g-k, SFig.2g-k). Fibers under compression ($\lambda < 1$) are oriented close to the angular direction, forming loops around the cell (Fig.3g-k). Within each of these loops, and close to the cell boundary, the stretch is nearly uniform for Family-1 models (Fig.3g,h). Similar behavior is seen in \cite[Fig.6]{Goren2020}.

Simulations with Family 2 exhibit two differences: the distribution of compressive stretch around the cell is strongly inhomogeneous (Fig.3k), and the maximum compression is up to twice as high as in Family-1 simulations, $60\%$ compressive strain (or stretch $\lambda$ $\approx$ 0.4) compared to $30\%$ ($\lambda$ $\approx$ 0.7) for Family 1 (colorbars in Fig.3g-k and SFig.2g-k). Compressed Family-2 fibers are still roughly in the angular direction. The most compressed fibers occur within narrow bands emanating radially from the cell and reaching as far as 6 deformed cell radii into the matrix (Fig.3k,m, SFig.2k,n). Furthermore, in Family 2, network triangles comprising the densified bands are excessively compressed (Fig.3m), as they contain fibers that have nearly collapsed. Fibers under tension are aligned along the axis of densification bands, roughly perpendicular to fibers under compression (Fig.3n). When the densification ratio of the networks in Fig.3c is compared to the compressed fiber distribution of Fig.3k, it becomes clear that regions of localized excessive densification ($\varrho \approx 3$) coincide with the bands containing severely compressed fibers (Fig.3c,k-n, SFig.2c,k).

In Family-1 simulations, severe compressive stretch is not observed at $50\%$ contraction level, with $\lambda$ remaining above 0.7, compared to 0.4 for Family 2. Densified zones are much shorter and confined to the immediate vicinity of the cell (Fig.3a,b, SFig.2a,b) with triangles less compressed ($\varrho$ at most 2 compared to 3 for Family-2).

\begin{figure}[H]
    \includegraphics[width=13cm]{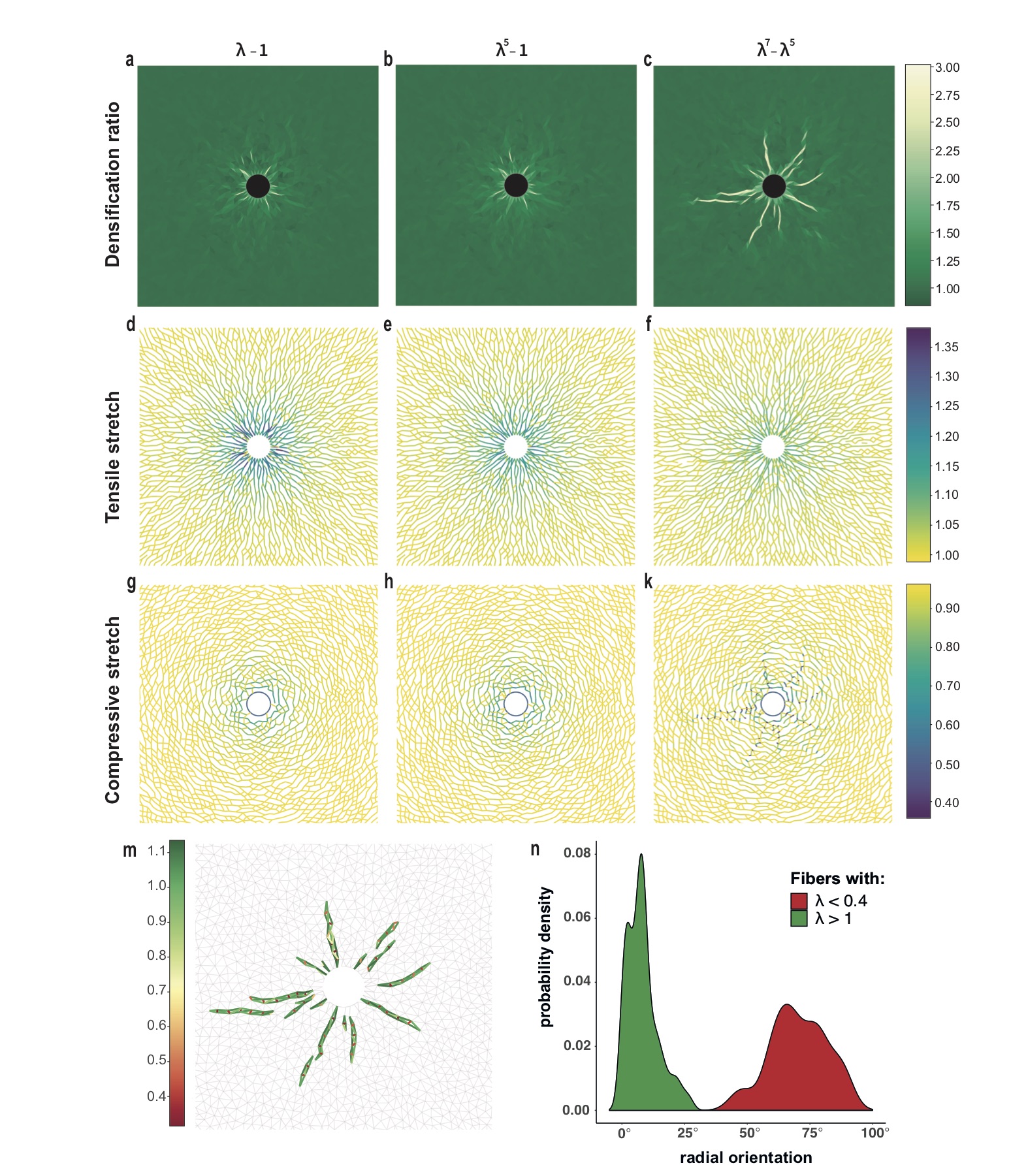}
    \label{res:fig1}
    \caption{\textbf{Fiber collapse instability and severe localized densification.} Simulations with a single cell at $50\%$ contraction with Family-1 models $\lambda - 1$ and $\lambda^{5} - 1$ and Family-2 model  $\lambda^{7} - \lambda^{5}$ \textbf{(a-c)} Densification ratio of triangular elements (color plot) in deformed networks \textbf{(d-f)} tensile stretches and  \textbf{(g-k)} compressive stretches in deformed fibers \textbf{(m)} stretch of deformed fibers and \textbf{(n)} radial orientation distribution of fibers within the densified bands in Family-2 case (c). \\Colorbars: (a-c) densification ratio $\rho$ of the deformed networks, (d-k) fiber stretch $\lambda$.}
\end{figure}

\subsubsection{Comparing critical levels for densification pattern formation in the two model families}
\label{results:single2}
Since the main difference between the two model families is the unstable compression regime in Family-2 models, where the $S(\lambda)$ curve has negative slope (Fig.1d), the results of \ref{results:single1} suggest that fiber collapse instability is responsible for the sudden growth of localized densification bands. In order to test this hypothesis, we simulate network response under gradual cell contraction, to study the onset of densified band formation.

Fig.4a-e shows a Family-2 network with a cell contracting at five consecutive levels, from $20\%$ to $40\%$ reduction of cell initial radius. Initially, as contraction progresses, the densification ratio of essentially the same few triangles proximal to the cell increases linearly with cell contraction (Fig.4,a-d) up to $35\%$. Remarkably, at the next level of ($40\%$) contraction, densified bands around the contracting cell have appeared, extending noticeably further into the matrix (Fig.4e). Below each plot of Fig.4a-e, in a "tree diagram", we plot the stretch of each individual fiber  (abscissa) versus distance from the cell center (ordinate) for each contraction level; color indicates orientation relative to the radial direction. The evident asymmetry near the base of each tree at larger contractions shows the difference of compressive versus tensile stretches. Tensile stretches $\lambda > 1$ grow gradually with increasing contraction. In fibers under compression ($\lambda < 1$), the stretch first decreases slowly, with only a few fibers in the unstable regime $\lambda < 0.85$ (red dotted line), all of whom are close to the cell up to $30\%$ contraction. At $35\%$ there is a steep increase in the number of fibers below the threshold, with stretches down to 0.4 and reaching more than 6 cell radii into the network by $40\%$ contraction. Going back to the respective densification ratio configurations, we observe that the jump in fiber compressive stretch and the abrupt appearance of densified bands occur at the same contraction level between $35\%$ and $40\%$.

This trend in densification localization is reflected in plots of the maximum over the network of the densification ratio inverse $1/\varrho_{max}$, and the minimum stretch $\lambda_{min}$,  at each contraction level (Fig.4f). We note that $1/\varrho_{max} = J_{min}$, the minimum area ratio, corresponding to the most compressed triangular element. The densification ratio first increases slowly with contraction, then there is a steep rise between 35 and $40\%$ contraction, the level at which extensive localized densification is spotted (Fig.4e). The minimum fiber stretch follows exactly the same behavior as $1/\varrho_{max}$, the two curves in Fig.4f being nearly identical. Initially, $\lambda_{min}$ decreases approximately linearly with contraction, namely the minimum stretch occurs at the cell boundary as dictated by the boundary conditions
\begin{equation}
    \label{lmin}
    \lambda_{min} = 1 - \gamma,
\end{equation}
with $\gamma$ the fractional cell diameter decrease. Then there is a sudden drop in stretch magnitude at $35 - 40\%$ contraction, exactly the level of sudden $1/\varrho_{max}$ drop in Fig.4f and band growth in Fig.4e. This shows that element densification is driven by fiber collapse as explained in Fig.2e. We recall also Fig.3m showing a collapsed red fiber within each densified green triangle (See Methods 2.2, \textit{Fiber collapse instability})

The behaviour of Family-1 networks is different (Fig.4g, SFig.3, SFig.4). The minimal stretch $\lambda_{min}$ follows \eqref{lmin} all the way up to the largest simulated contraction (red line in Fig.4g), occurs on the cell boundary, and is equal to cell boundary contraction prescribed by boundary conditions. This shows that fiber collapse is not observed, as expected. In contrast, the maximal densification ratio does undergo a sudden leap ($1/\varrho_{max}$ drop in Fig.4g) as in Family-2 models, albeit at a higher contraction level of $\gamma \approx 45\% - 50\%$. This is evidence of an element collapse instability (See Methods 2.2) that is weaker and requires higher cell contraction than the fiber collapse instability of Family-2 models. Notably, this instability occurs in the linear fiber model $S(\lambda) = \lambda -1$ (SFig.3) as well as the nonlinear ones (SFig.4). 

\begin{figure}[H]
    \includegraphics[width=13cm]{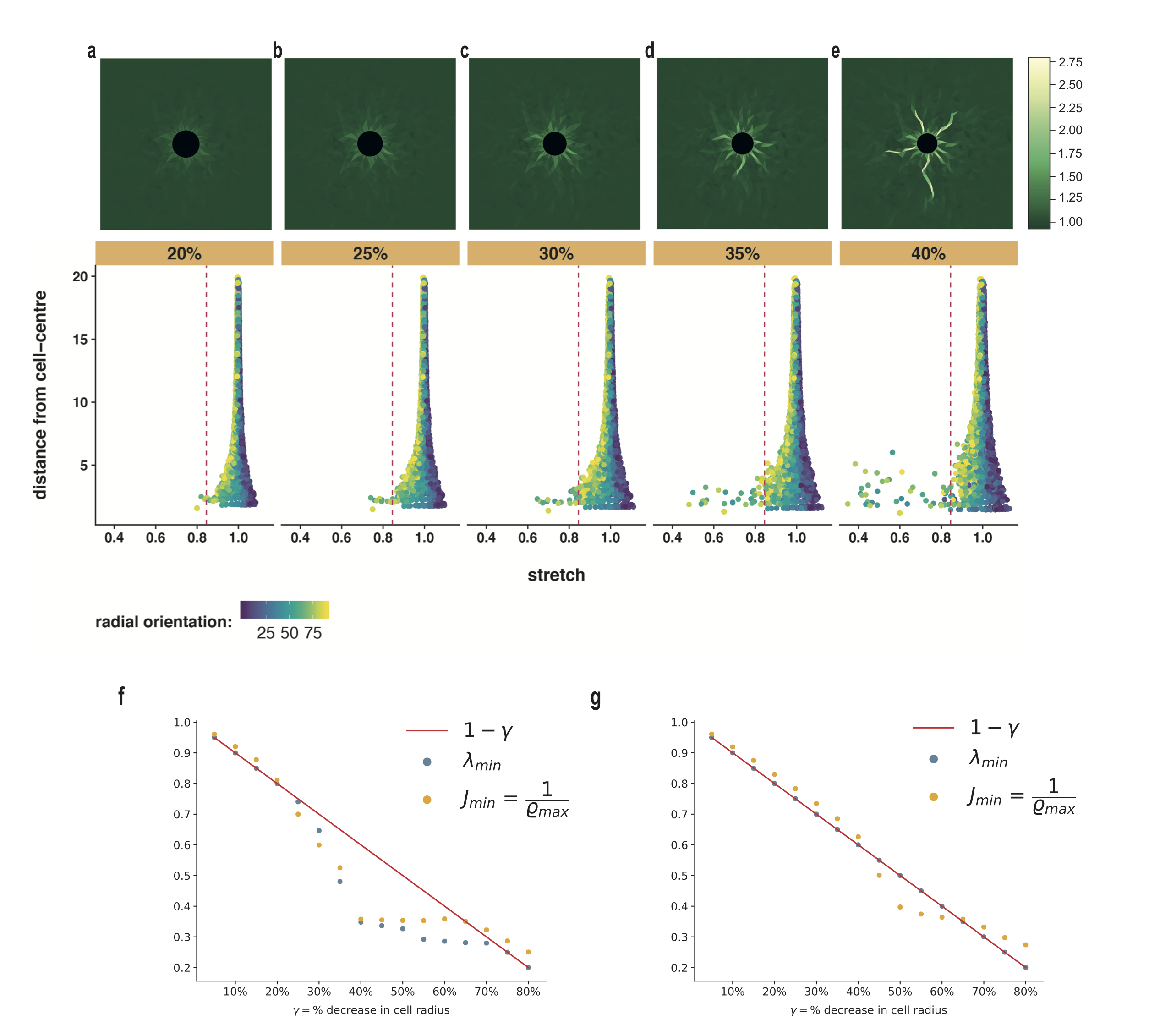}
    \label{res:fig2}
    \caption{\textbf{Progressive cell contraction and densification localization.} \textbf{(a-e)} Simulations with Family-2 model $S(\lambda) = \lambda^{7} - \lambda^{5}$ of a cell contracting in the range $5\% - 80 \%$. Top: densification ratio $\varrho$ color plot at each indicated contraction step. Bottom: tree diagrams,
    fiber distance from cell center versus fiber stretch for all fibers in the network at each  contraction step, \textit{x axis}: fiber stretch, \textit{y axis:} fiber distance from cell center.
    \textbf{(f,g)} Maximum densification ratio inverse $1/\varrho_{max}$ and minimum stretch value $\lambda_{min}$ over the network at each contraction level, for Family-2 $S(\lambda) = \lambda^{7} - \lambda^{5}$ and Family-1 $S(\lambda) = \lambda - 1$ respectively. Note that $1/\varrho_{max} = J_{min}$. Red solid line: cell boundary stretch imposed by boundary conditions. 
    \\ Colorbars: Densification ratio $\varrho$ and fiber radial orientation (in degrees). }
\end{figure}

\subsection{Intercellular tether formation in two-cell simulations}
We report on simulations involving a pair of cells contracting at $50\%$ of their initial radius, separated by either $6r_{c}$ or $4r_{c}$, where $r_{c}$ is the cell radius (Fig.5, SFig.5, SFig.6). What distinguishes these from singe-cell simulations is the spontaneous appearance of intercellular tethers, composed of thin, roughly parallel bands of high densification and fiber alignment, that connect the two cells (Fig.5c, SFig.5c). When cells are separated by a larger distance, $6r_{c}$, tethers are generated only with Family-2 models. Additional densified bands emanate radially from each cell (Fig.5c, SFig.5c) as before. In contrast, in Family-1 simulations, matrix densification is limited close to the cell boundary and cells remain isolated and disconnected (Fig.5a,b, SFig.5a,b). When cells are closer together, tethers are generated by all models, even  the linear one (SFig.6). In this case, we observe that they are substantially stronger in Family-2 simulations, as they extend from one cell to the other and are noticeably wider compared to Family-1 tethers (SFig.6a-c).

When tethers form, we observe a fraction of fibers, located almost entirely in the intercellular region, to be highly stretched (SFig.6d-f). These fibers are densely packed and aligned with the horizontal direction passing through the cell centers, generating straight paths of fibers connecting the two cells. These paths comprise the tether. At the same time, fibers under extreme compression occupy the same region as the tensile ones, but their orientation is nearly perpendicular to the paths of the tensed and aligned fibers (SFig.6g-k). This is true for all models, though fiber compression magnitude is almost twice as large with Family 2, reaching approximately $70\%$ compression (SFig.6k). This indicates that in Family-2 tethers, compressed fibers are well within the regime of the fiber collapse instability. In addition, we observe highly compressed fibers within the densified bands that emanate radially from the cell periphery (SFig.6k). 

When cells are separated by a greater distance, $6r_{c}$, and tethers are generated only with Family-2 models, fiber stretches highlight a significant difference between families (Fig.5d-k, SFig.5d-k). In Family-2 simulations, fiber distributions and orientations within the tether are the same as for shorter distances (Fig.5f,m, SFig.5f,k). For Family-1 models, this is no longer true, as the fiber paths are disrupted and tensile stretches are distributed in a broader region between cells, without the strong alignment we have with Family-2 models (Fig.5d,e, SFig.5d,e). This is reflected in angle distributions of the tensile fiber orientation, which are substantially different across models within the intercellular region (SFig.5m). This distribution is more localised for Family-2 models, consistent with greater alignment.

Excessively tensed fiber angles within the densified region are narrowly distributed about zero (horizontal direction through cell centers) (Fig.5m). Compressed fibers are distributed about $80-90 ^{\circ}$ within the densified region, compared to a uniform distribution in non-densified regions (Fig.5m). On the contrary, compressive stretches in Family-1 models are confined to concentric loops around each individual cell  instead of the region between cells, and oriented in the circumferential direction (Fig.5g,h, SFig.5g,h).

The previous findings hold for $50 \%$ contraction. When cells contract more,  tethers are eventually generated for Family-1 models as well. In Fig.6 we present the case of Family-1 model $S(\lambda) = \lambda^{5} - 1$ (eq. \eqref{Fam1}, Fig.1c) with two cells separated by $6r_{c}$ at four contraction levels $45 \%, 55 \%, 65\%$ and $75\%$. We observe that densification between the two cells progressively strengthens. Fiber compression in the cell-cell vicinity is ever-increasing with contraction level, resulting finally in a solid tether at $75\%$ contraction (Fig.6d). 

Working in the same manner for each model separately, we have tested different contraction levels ranging from $5\%$ to $80\%$ decrease in cell radii, for multiple distances separating the two cells. Results are summarized in Fig.6e. In particular, for each model we obtain a curve that indicates the minimum contraction cells should undergo to produce a tether, expressed as a function of cell distance. That is, above each curve a tether is predicted to form for the respective model. Clearly, Family-2 models are able to sustain tether formation for moderate contraction levels $\leq 50 \%$, and for relatively large cell-cell distances (up to $11r_{c}$). On the contrary, regarding Family-1 models for the same contraction levels $\leq 50 \%$, a full tether is formed when cells have a distance at most $5r_{c}$. 
 
What are the mechanisms responsible for the differences between tethers in the two Family models? Extremely compressed fibers occur in Family-2 tethers ($\lambda_{min} \approx 0.3$) but not in Family 1, where $\lambda_{min} \approx 0.6$ (Fig.7a,b). In Family 2, fiber collapse (extreme fiber compression, sudden $\lambda_{min}$ drop in Fig.7b) occurs at the same time as extreme densification (sudden $1/\varrho$ drop , Fig.7b). On the contrary, in Family 1, we see extreme densification without fiber collapse (Fig.7a). In Family 2, most collapsed triangles within the tether contain a highly compressed  fiber, oriented within $45 \degree$ of the vertical (Fig.7d as in Fig2e). In contrast, in Family-1, collapsed triangles have nearly horizontal bases, while  the other two sides are under moderate compression, and are closer to horizontal than vertical after collapse (Fig.7c as in Fig.2f). These findings indicate that fiber collapse instability is the main player in Family-2 tethers, whereas the dominant role in Family-1 tethers is played by element (triangle) collapse instability.

\begin{figure}[H]
    \includegraphics[width=13cm]{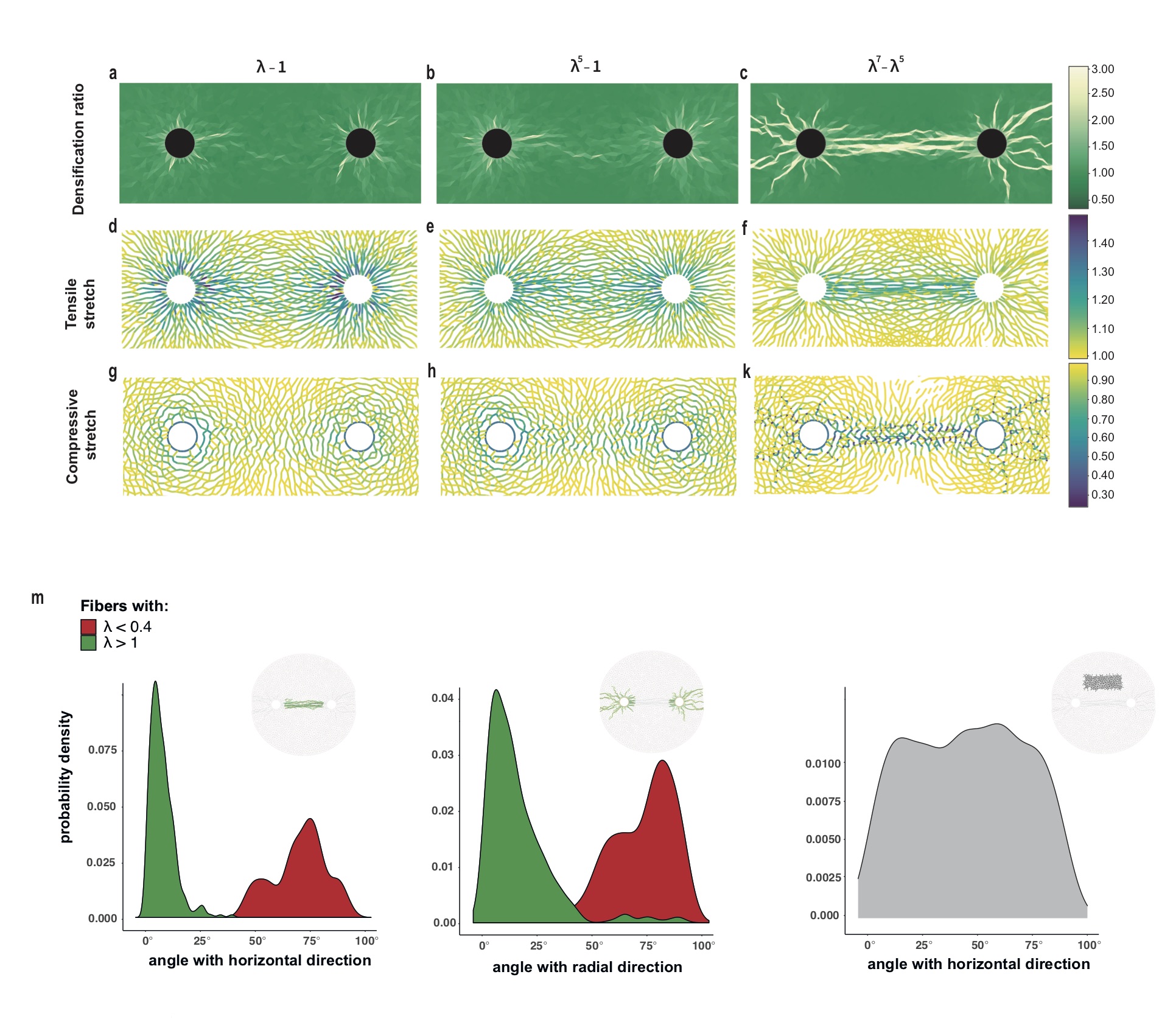}
    \label{res:fig3}
    \caption{\textbf{Intercellular tether formation.} Simulations with two cells contracting at $50\%$, for three different models (three columns in \textbf{a-k}). Cell centers are separated by $6r_{c}$, where $r_{c}$ is the undeformed cell radius. \textbf{(a-c)} densification ratio of triangular elements (color plot) in deformed networks \textbf{(d-f)} tensile stretches and \textbf{(g-k)} compressive stretches of deformed fibers \textbf{(m)} orientation distribution of deformed fibers within the densified zones (tether and radial bands) in Family-2 case (c) and within the highlighted non-densified zones. Horizontal direction is the one parallel to the axis connecting the cell centres. Radial direction is the one passing through the cell centre. \\Colorbars: (a-c) densification ratio $\varrho$ of the deformed networks, (d-k) fiber stretch.}
\end{figure}

\begin{figure}[H]
    \includegraphics[width=15cm]{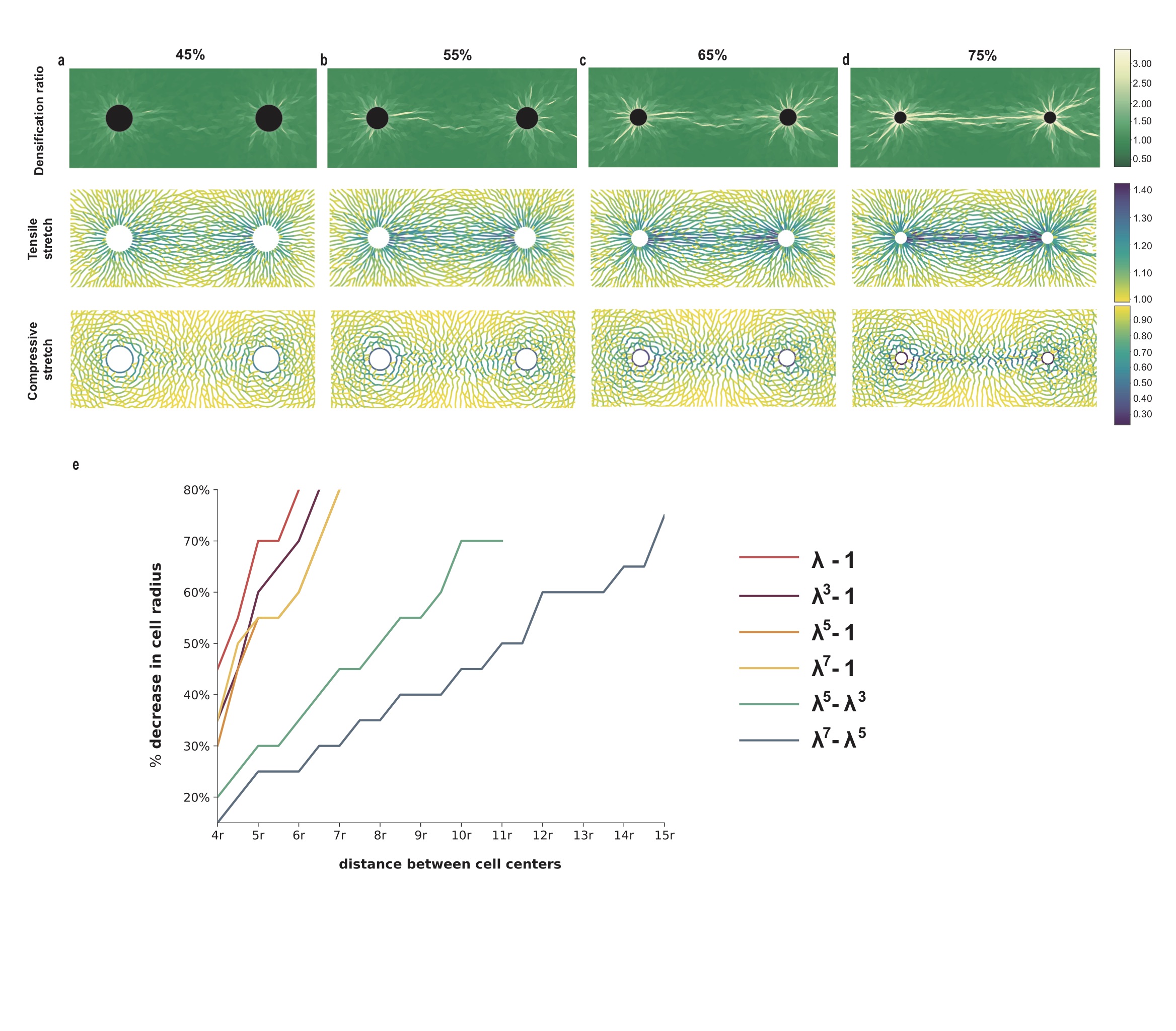}
    \label{res:fig4}
    \caption{ \textbf{(a-d)} \textbf{Tether formation in Family-1 networks.} Simulations with two cells contracting at different levels in a Family-1 network (model $S(\lambda) = \lambda^{5} - 1$). Cell centers have distance $6r_{c}$, where $r_{c}$ is the undeformed cell radius. Densification ratio color plot of triangular elements (up), tensile (middle) and compressive (bottom) stretches of deformed fibers at each contraction step. \textbf{(e)} \textbf{Contraction versus cell-cell distance required for tether formation in various models.} 
    Simulations with two cells contracting in the range  $5\%-80\%$ decrease in cell radius (\textit{y axis}). Cells are separated by a distance proportional to cell undeformed radius $r$ (\textit{x axis}). Each curve  corresponds to a different model and depicts the minimum contraction level required for a solid tether joining the cells as a function of distance separating them.}
\end{figure}

\begin{figure}[H]
   \hspace{-0.5cm} \includegraphics[width=16cm]{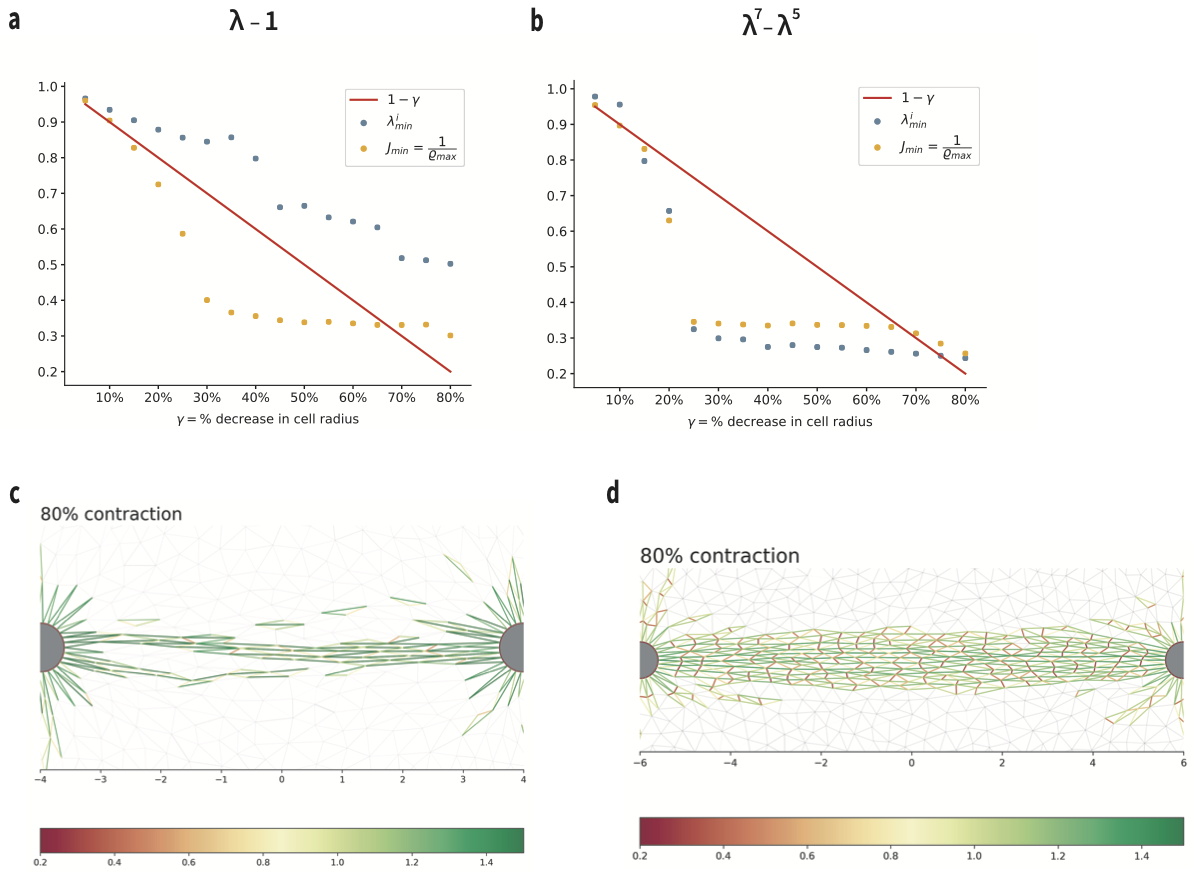}
    \label{res:fig7}
    \caption{\textbf{Mechanisms of densification within tethers in the two Families.} \textbf{(a-b)} Maximum densification ratio inverse $1/\varrho_{max}$ and minimum stretch value $\lambda^{i}_{min}$ over the network at each contraction level excluding fibers on the cell boundary, for Family-1: $S(\lambda) = \lambda - 1$ and Family-2: $S(\lambda) = \lambda^{7} - \lambda^{5}$  respectively. Note that $1/\varrho_{max} = J_{min}$. Red solid line: cell boundary stretch imposed by boundary conditions.
    \textbf{(a)} Element area collapse (yellow dots) occurs without fiber collapse (blue dots), indicating element collapse instability. \textbf{(b)} Densification (yellow dots) occurs simultaneously with fiber collapse (blue dots), suggesting fiber collapse instability.
    \textbf{(c-d)} Stretch of fibers (red: compression; green: tension) at $80 \%$ contraction within a Family-1 tether (c) and Family-2 tether (d). Note scarcity of compressed fibers despite presence of collapsed triangles in (c), indicating element collapse instability. In contrast, in (d) most collapsed triangles contain a highly compressed fiber, pointing towards fiber collapse instability.}
\end{figure}

\newpage
\section{Discussion}
Extracellular Matrix (ECM) mechanical remodelling by cellular forces brings about unique deformation patterns of excessive matrix densification and fiber alignment, both playing a central role in intercellular communication \cite{Harris1981,Stopak1982, Vader2009,Notbohm2015} and in cell motility  and invasion \cite{Provenzano2008, Ferruzzi2019}. In order to explore this type of cell-induced ECM deformations, we develop a discrete model that accounts for individual fibers and their intrinsic mechanics. The discrete fiber network model featured in this study complements our prior work \cite{Grekas2021} where we demonstrated through continuum modelling and experiments that the aforementioned phenomena are the result of a material instability. The latter is caused by a special nonlinearity due to fiber buckling under compression. To understand how discreteness affects these results, here we implement two different families of fiber constitutive relations, with distinct nonlinearity and stability features. Family 1 displays a positive but decreasing stiffness with increasing compression, while Family 2 entails a stretch instability phase, where stiffness becomes negative at extreme compressions. Family 1 represents the traditional view of post-buckling behavior. Family 2 is a more radical model, incorporating recent experimental observations on buckling of hierarchical beams \cite{Tarantino2019}. 

Our simulations results have revealed instability mechanisms that have not been identified in previous work. In particular, all nonlinear spring networks are susceptible to element collapse (triangle buckling) instability; see Methods. This  includes linear spring networks provided large rotations are accounted for \cite{pecknold1985}. In Family-2 networks,  an additional fiber collapse instability occurs because of  total loss of strength due to buckling of  fibers in compression (Methods). Simple geometry shows that fiber collapse implies element collapse (Fig.2e), but not vice versa. Thus both instabilities can occur in Family-2 networks. In contrast, Family-1 networks can undergo element collapse, despite the fact that individual fibers exhibit stable behavior. The distinction of these different instabilities is an effect of discreteness and is not captured by continuum models, including ones that allow instability \cite{Grekas2021}.
    
In Family-2 models we observe a sudden increase in densification simultaneously with abrupt fiber collapse, both in single and two-cell simulations (Fig.4f and Fig.7b). The majority of the elements in the densified regions contains a severely compressed fiber (red fibers in Fig.7d). This is strong evidence that the mechanism behind densification in Family-2 networks is fiber collapse instability (Methods and Fig.2e), driven by cell compression. Densification also occurs in Family-1 networks, but requires  higher levels of cell compression. Fiber collapse is not encountered in this case (Fig.4g and Fig.7a). Instead, densification is due to triangular elements collapsing (Fig.7c). Element collapse instability (Methods and Fig.2f) is the main player behind the appearance of densified zones in Family-1 networks. These observations apply both to intercellular tethers and densified zones around single cells.
    
One important finding concerns fiber alignment. Regardless of the model, simulations show that fiber excessive alignment occurs simultaneously with densification and at the same locations in the ECM, both in single and two-cell cases (Fig.3, Fig.5). In particular, in Family 1 we see that severe element compression forces all sides of the densified triangles to align with each other (Fig.7c), a direct effect of element collapse instability (Fig.2f). Especially in two-cell cases, before tethers form we observe a moderate tendency of fibers to align (Fig.6a,b). After tethers form, fibers within the tethers are aligned almost perfectly (Fig.6d) and the elements they belong to are the ones that show extreme densification. In Family-2 tethers and bands, we see the stretched sides of the densified elements aligned with each other while the highly compressed one is roughly perpendicular to them (Fig.3m, Fig.7d), evidence that fiber collapse instability brings about the alignment of stretched fibers within the densified regions. These significant observations indicate that ECM compression instabilities are responsible for both matrix densification and fiber alignment.
    
In general it is much easier to form a tether with Family-2 models than with Family-1. For the same distance between two cells, a Family-1 tether requires a much higher compression in order to form. For example, for a distance $6r_{c}$, where $r_{c}$ is cell radius, the linear model requires $80\%$ compression whereas $25\%$ is sufficient with $\lambda^{7} - \lambda^{5}$ (Fig.6e). Given a level of compression, say $50\%$, a Family-1 tether is formed when cells are very close, less than $5r_{c}$. On the contrary, a Family-2 tether can form when cells are more than twice as far away from  each other (Fig.6e).

Our most prominent prediction is the formation of densified tethers between two cells, and densified radial bands emanating from each cell. Zones of high densification (tethers) have been observed experimentally joining two clusters of contractile cells \cite{Harris1981} \cite{Stopak1982} \cite{Shi2014} \cite{Ban2018} while thinner bands were seen to emerge from each cluster \cite{Harris1981} \cite{Stopak1982}, extending and gradually diminishing within the matrix. Densified tethers and radial bands were also observed in \cite{Grekas2021} where contracting active particles were used in place of living cells, thereby excluding non-mechanical causes behind densification. Fibers within the tethers are highly aligned along the tether axis. Individual cells from each explant \cite{Harris1981} \cite{Stopak1982} or acinus \cite{Shi2014}  start migrating along the tether in an attempt to reach the neighbouring cluster. Moreover, isolated fibroblasts grew protrusions towards each other, along the tether that formed following their contraction \cite{Notbohm2015}. These studies illustrate the significance of tethers and radial bands in cell migration, motility and intercellular communication. Our simulations identify the formation of these densified zones as a direct consequence of compression instabilities. An additional effect of these instabilities is the close alignment of fibers within the densified zones. Alignment and densification are therefore seen to be part of the same mechanism. 

The most essential application of this work is cancer invasion and metastasis. Tumor explants cultured in an initially randomly organized matrix aligned the collagen fibers around them by contracting. This allowed individual cancer cells to use the tracts of aligned fibers as highway paths to invade the ECM \cite{Provenzano2006}  \cite{Provenzano2008}. In fact, both densification \cite{Provenzano2008b} and fiber alignment \cite{Provenzano2006} are considered prognostic biomarkers for breast carcinoma \cite{Provenzano2011}, specifically, ``bundles of straightened and aligned collagen fibers that are oriented perpendicular to the tumor boundary''   \cite{Provenzano2011}. Contractility is a necessary ingredient in their formation \cite{Provenzano2008}. All these observations apply to both tethers and radial densified bands. A different phenomenon is seen in expanding tumors where the densified layer surrounding them consists of fibers parallel (not perpendicular) to their boundary \cite{Provenzano2011}. Remarkably, our simulations of an expanding cell confirm this, see Supplementary Material, SFig.7. Additionally, aligned collagen fibers offer biochemical molecules transportation between cells \cite{Gomez2019}. Focusing on the elevated fiber alignment within the tethers, our predictions highlight the contribution of compression instability to cancer related ECM mechanisms.

Family-2 models (unstable stretch responses) result in well-defined tethers with highly localised densification and fiber alignment very close to the tether axis. These results are in qualitative agreement with experiments \cite{Stopak1982, Shi2014, Grekas2021}. On the contrary, in models with stable stretch response (Family 1) tethers are diffused and not localized, while fibers under tension distribute in the broader intercellular region. In addition, tethers with Family 2 occur under experimentally observed physiological levels of cell contraction $\approx 50\%$ \cite{Shi2014, Notbohm2015, Grekas2021}, in contrast to Family 1 where extreme contraction is required. Considering the above, Family-2 models are preferable over Family-1 as they describe  experimental observations better. The unstable response of Family-2 fibers is justified by recent work \cite{Tarantino2019} on post-buckling behavior of beams with hierarchical structure, such as ECM fibers \cite{Piechocka2010} (see Methods). 
    
There have been considerable efforts in ECM modelling \cite{Notbohm2015, Rosakis2015, Abhilash2014, Ronceray2015, Xu2015, Liang2016, Grimmer2018, Sopher2018, Goren2020, Mann2019}. In these studies, ECM fibers are usually modelled as Timoshenko beams  \cite{Abhilash2014, Grimmer2018} or as elements with asymmetric elastic responses to extension and compression, obeying either piecewise linear stress-strain curves or combining strain-stiffening with compression softening intended to model fiber buckling \cite{Notbohm2015, Rosakis2015, Xu2015, Sopher2018, Goren2020, Liang2016, Mann2019}. Even though these approaches have explored nonlinear aspects of fiber behavior, they are limited by stable stretch responses (monotonic) and sometimes small deformations. As far as the stretch response is concerned, all these models are similar in spirit to our Family-1 models. None of these works address instability. Here, we recognise that instability plays a central role in the appearance of ECM densification and fiber alignment. Our work shows that instability can occur even in models with stable stretch response (Family 1), but it does so at unreasonably high cell contraction levels. This prompts us to introduce models (Family 2) whose stretch response becomes unstable in compression. This particular instability allows tether formation and fiber alignment under experimentally observed cell contraction levels. 

Fiber alignment has been explored by previous studies \cite{Sopher2018},\cite{Goren2020}, yet as a separate ECM mechanism involved together with compression buckling in intercellular force transmission \cite{Sopher2018} or matrix elastic anisotropy \cite{Goren2020}. Sopher et al., \cite{Sopher2018} suggest that elevated tension in the intercellular region obliges fibers to stretch and align.
In our simulations densification and fiber alignment suddenly jump to much higher values compared to aforementioned works. This transition occurs when a compression instability is activated. For models with stable stretch response (Family 1), this would occur in much higher contraction levels than considered in previous works \cite{Notbohm2015, Rosakis2015, Xu2015, Sopher2018, Goren2020, Liang2016, Mann2019}, which explains the lower levels of densification and alignment observed there. Family-2 models not only require moderate cell contraction, but also densification is much stronger, tethers are solid and substantially wider and fibers almost perfectly aligned to the tether axis.

Deformation induced anisotropy has been studied \cite{Goren2020} as a possible mechanism for long-range cell communication. We remark that the compressive instabilities studied here create strong anisotropy in the densified state of the network, because they create a strongly aligned and dense uniaxial distribution of fibers from an originally roughly isotropic random fiber distribution. Our results show that there is a single unifying mechanism behind densification, fiber alignment and matrix anisotropy, and this is compression instability. These phenomena occur simultaneously within the same localized zones as soon as compression instability is triggered either due to fiber buckling (collapse) or element collapse (snapthrough buckling). 

Compression instability due to buckling of elastic fibers in networks was first identified in \cite{Lakes1993} as a mechanism of localized densification. The continuum model of \cite{grekas2019, Grekas2021}, obtained from a fiber model through orientational averaging, also exhibits a compression instability, and predicts highly localized densified tethers and radial bands. Their morphology is qualitatively similar to the ones reported here. In general, it is practically impossible to obtain an explicit continuum constitutive law for a random network through a rigorous limit as the fiber length approaches zero \cite{Gloria2014,Friesecke2002}. The advantage of our model is that it captures the discrete nature of the actual fibrous network. As a result it can distinguish between different types of compression instabilities (buckling of fibers versus fiber elements) and it clarifies the close connection of instability with fiber alignment and densification. 

Possible extensions of our work include three-dimensional network models, more sophisticated modelling of joints and crosslinking between fibers, viscoelastic effects in fiber behavior, and the possibility of larger scale instabilities in analogy with periodic discrete networks \cite{Friesecke2002, muller1993}.

Our models highlight compression instability due to buckling as a crucial nonlinear mechanism underlying the mechanical behavior of fibrous ECM and give rise to new insights in exploring the nature of cell-induced deformations that underlie matrix densification and fiber alignment and their implications in intercellular biomechanical interaction, cancer metastasis and cell motility. 

\newpage

\noindent \textbf{Competing interests.}
We declare we have no competing interests.
\\
\textbf{Funding.} This work was financially supported by Fonds National de la Recherche (FNR), Luxembourg through AFR PhD programme (AFR2020/14582656) and under the PRIDE programme (PRIDE17/12252781, DTU DRIVEN). Georgios Grekas was partially supported by the Vannevar Bush postdoctoral fellowship.

\bibliographystyle{model1-num-names}
\bibliography{sample.bib}

\newpage
\beginsupplement

\section{Supplementary Material}

\begin{figure}[H]
\vspace{-0.5cm}
   \hspace{2cm} \includegraphics[width=9.5cm]{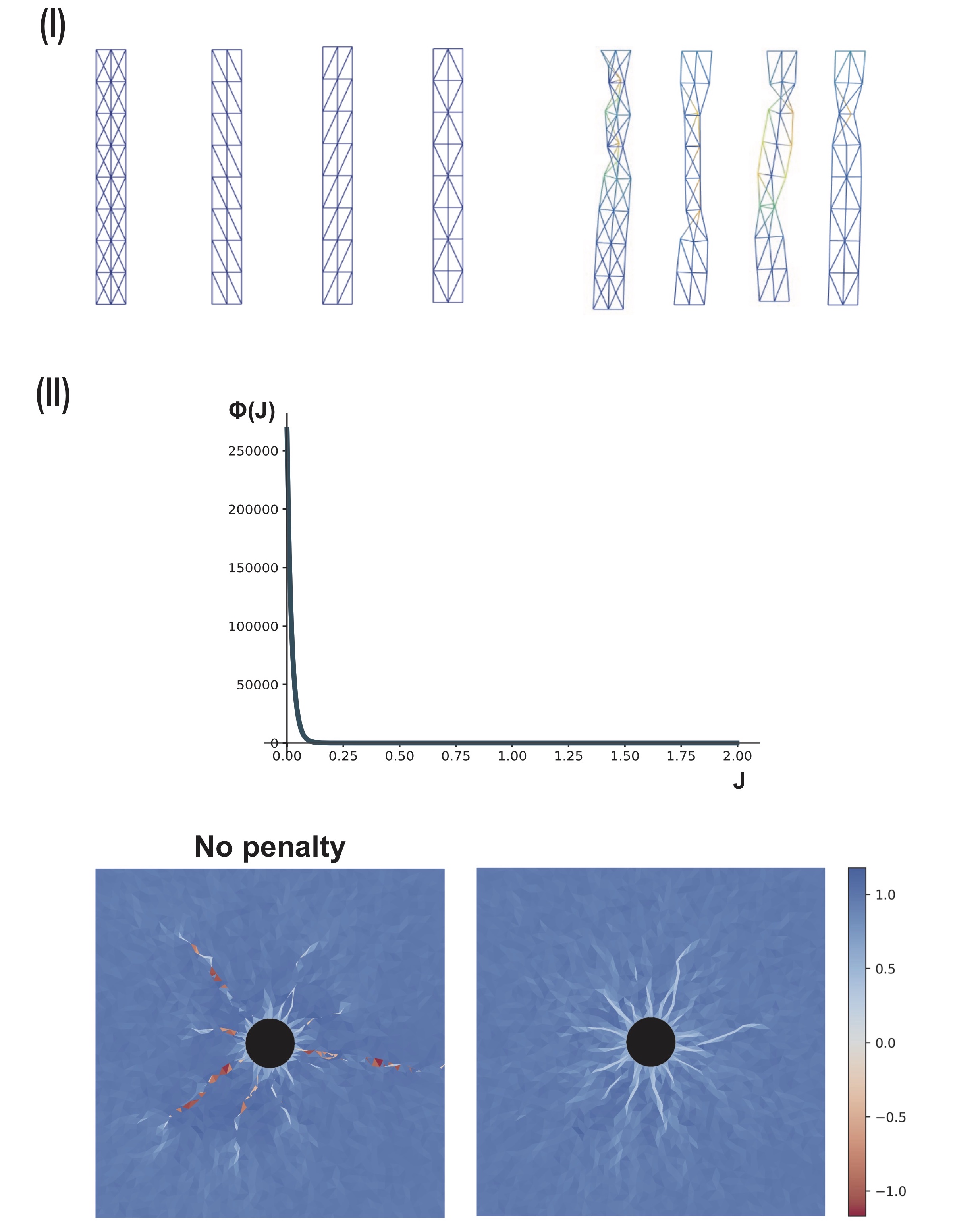}\label{sfig1}
    \caption{\textbf{Interpenetration of matter and penalty term.} \textbf{(I)} Various presentations of triangulated rectangular truss elements. Each edge in the structures represents a linear spring. Dirichlet boundary conditions were applied on the upper boundary nodes by imposing a displacement u = (\textit{h}, 0.0), \textit{h} being the scale to \textit{x} direction. Deformed structures contain triangles that have changed orientation, resulting in interpenetration of matter. \textbf{(II)} Top: Penalty term $\Phi(J)= exp(-Q(J-b))$, where $J$ is \textit{ratio of deformed to undeformed oriented triangle area}. $Q>0$ is large and $b>0$ is small constant. As a result, negative values of $J$ have high energy cost, whereas positive values have negligible contribution to the network's total energy. Bottom: Simulations of a cell contracting by $50\%$, either with or without the penalty term for the area ratio $J$. Without penalizing $J$, the optimizer finds solutions that are physically unacceptable, as $J<0$ corresponds to elements (red) that changed orientation. Colorbar: $J$ values.}
\end{figure}

\begin{figure}[H]
    \includegraphics[width=13cm]{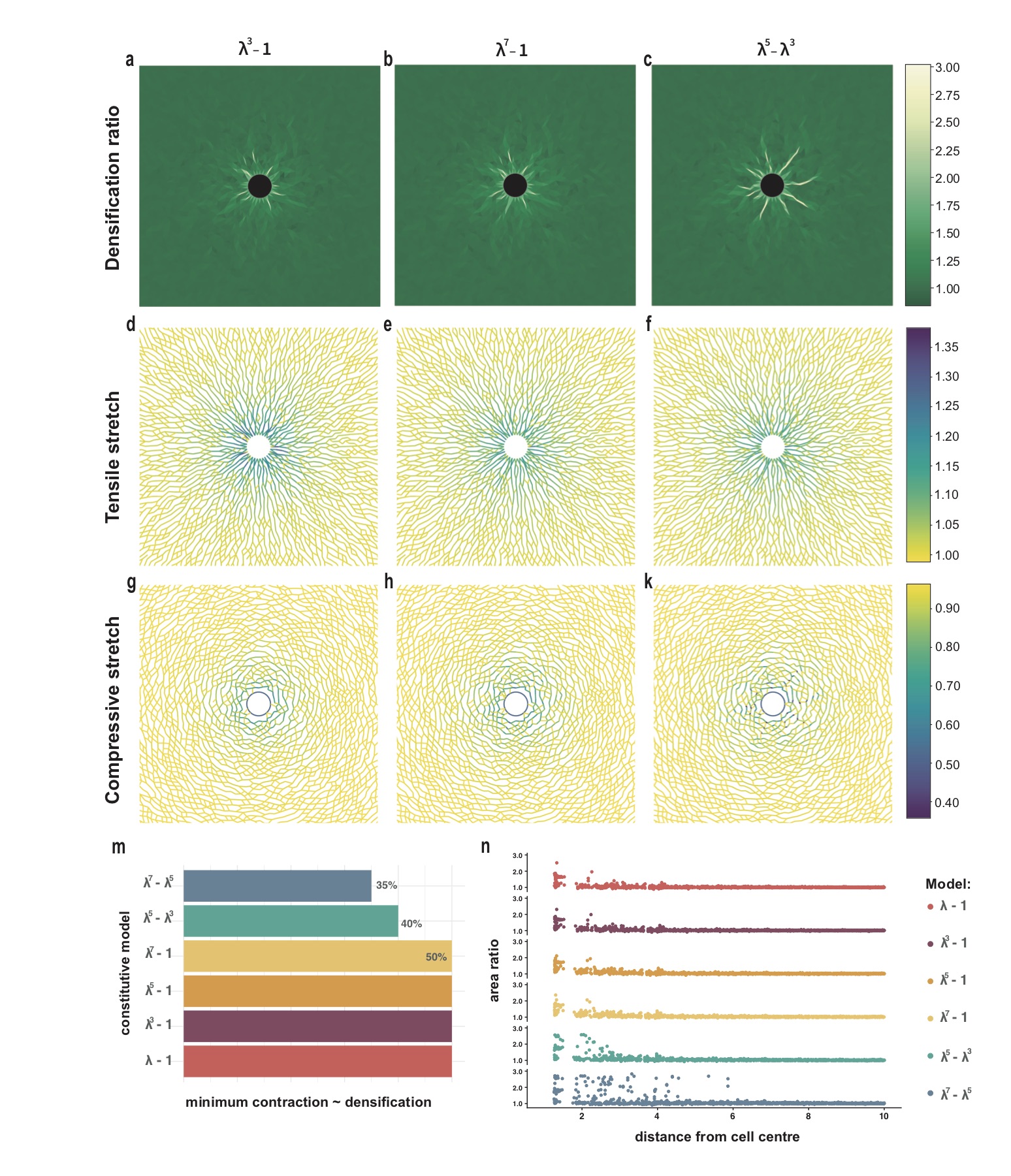}
    \label{sfig2}
    \caption{\textbf{Fiber collapse instability and severe localized densification.} Complementary to Results, Fig.3 containing simulations of a single cell at $50\%$ contraction with Family-1 models $\lambda^{3} - 1$ and $\lambda^{7} - 1$ and Family-2 model  $\lambda^{5} - \lambda^{3}$. \textbf{(a-c)} Densification ratio of triangular elements (color plot) in deformed networks \textbf{(d-f)} tensile stretches and \textbf{(g-k)} compressive stretches in deformed fibers. \textbf{(m)}  Minimum contraction required for densification to be evident for each one of the models studied. \textbf{(n)} Simulations with various models of one cell contracting at $50\%$; \textit{x axis:} triangular element distance from cell center, \textit{y axis:} element densification ratio.\\
    Colorbars: (a-c) densification ratio $\varrho$ of the deformed networks, (d-k) fiber stretch $\lambda$
}
\end{figure}

\begin{figure}[H]
    \includegraphics[width=13cm]{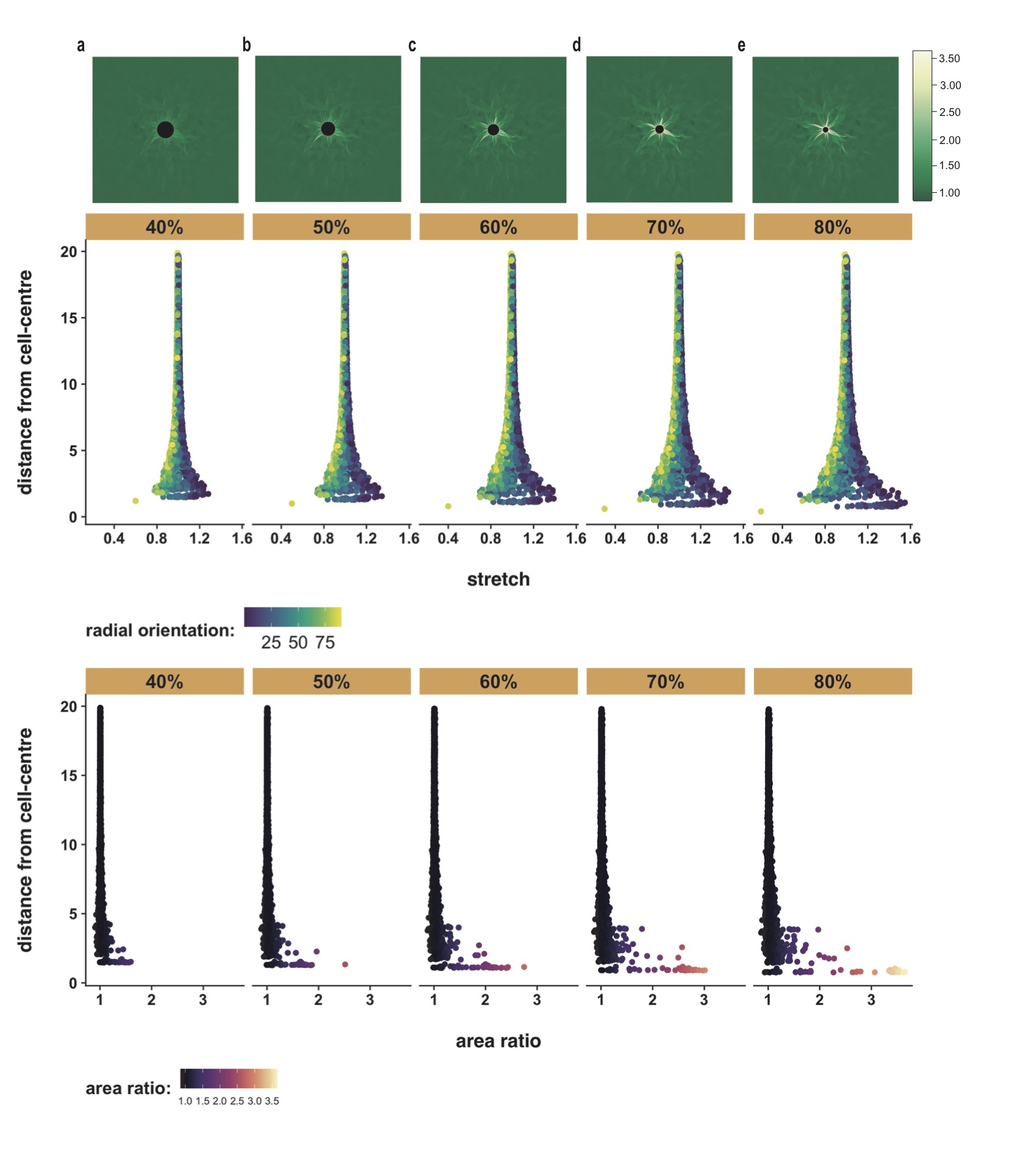}
    \label{sfig3}
    \caption{\textbf{Progressive cell contraction and matrix localization.} Complementary to Results, Fig.4. Simulations with the linear Family-1 model $S(\lambda) = \lambda - 1$ of a cell contracting in the range $5\% - 80\%$. \textit{Top:} densification ratio $\varrho$ color plot at each indicated contraction step. \textit{Middle:} tree diagrams, fiber distance from cell center versus fiber stretch for all fibers in the network at each contraction step, x axis: fiber stretch, y axis: fiber distance from cell center. \textit{Bottom:} triangular element distance from cell center versus densification area ratio, \textit{x axis:} densification area ratio $\varrho$, \textit{y axis:} triangular element distance from cell center.}
\end{figure}

\begin{figure}[H]
    \includegraphics[width=15cm]{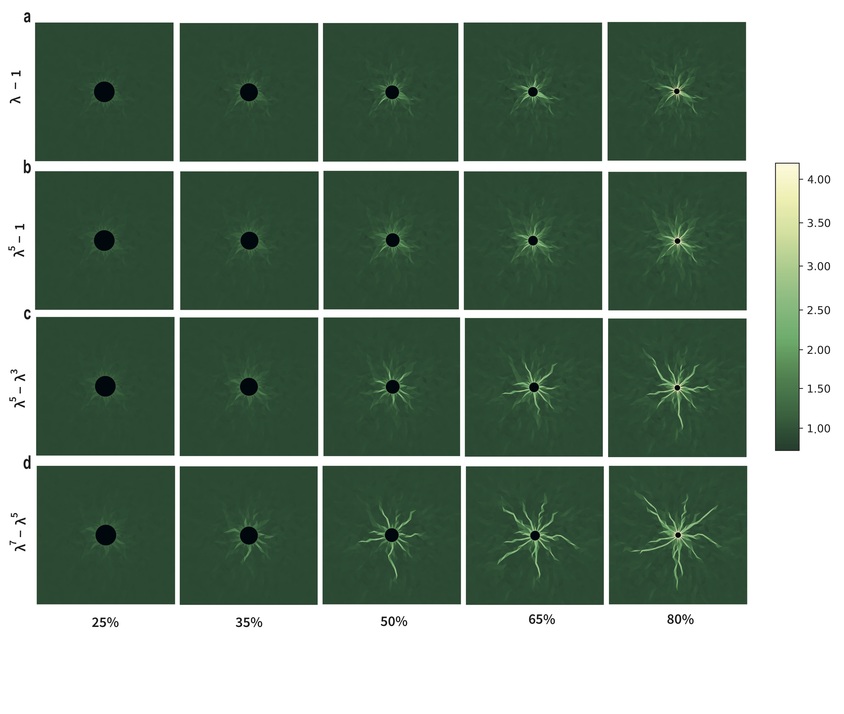}\label{sfig4}
    \caption{\textbf{Progressive contraction and densification.} \textbf{(a-b)} As contraction level rises, densification strengthens in the close proximity of the cell for Family-1 models. The bands consisting of densified elements do not propagate far from the cell boundary, reaching as far as 3 deformed cell radii at $80\%$. On the contrary, in Family-2 simulations \textbf{(c-d)} densification is evident at much lower contraction levels, $35\%$. With increased contraction, more densified bands are generated and extend substantially further into the matrix.
    Colorbar: densification ratio $\varrho$ of deformed networks.}
\end{figure}

\begin{figure}[H]
    \includegraphics[width=13cm]{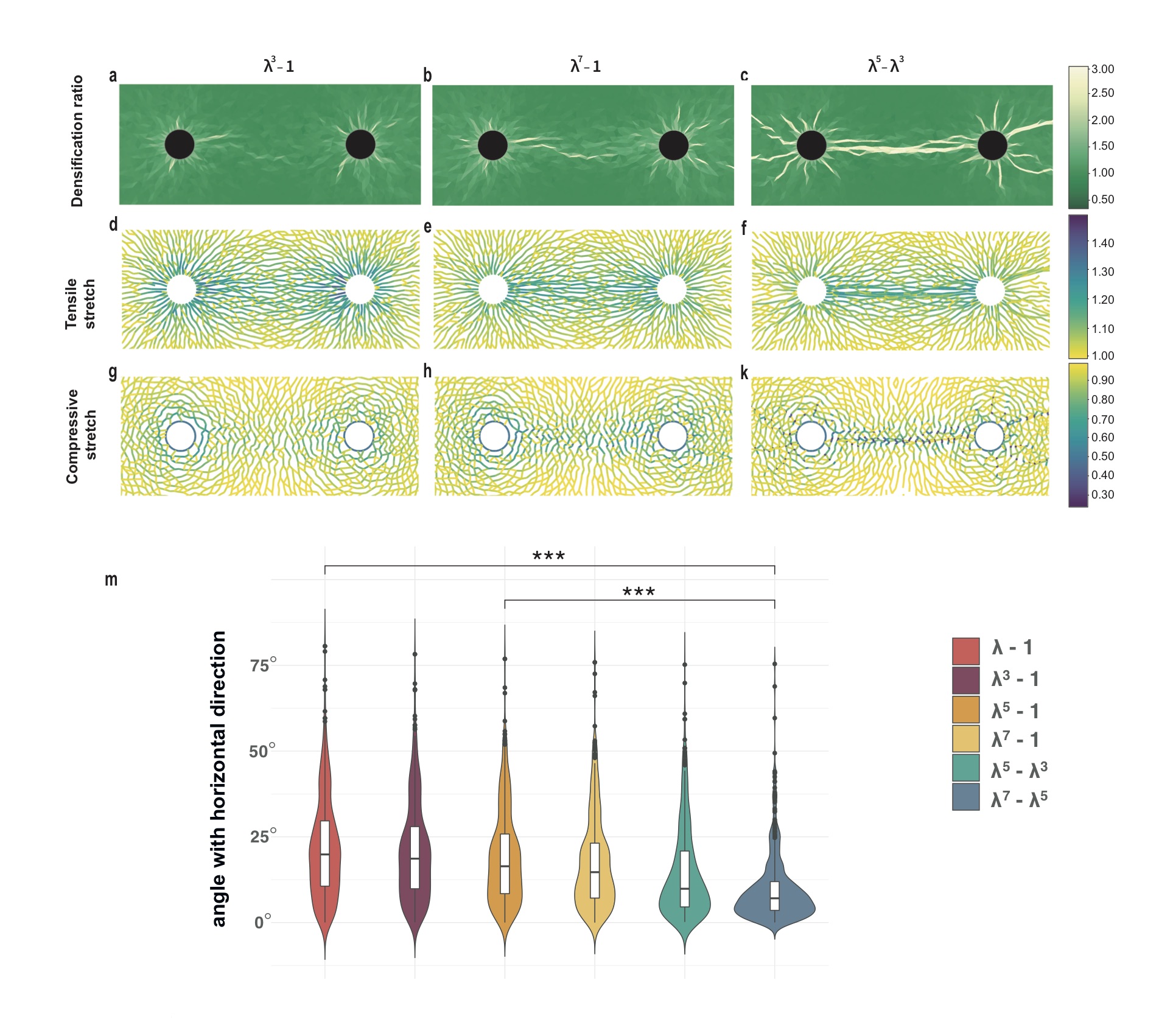}
    \label{sfig5}
    \caption{\textbf{Intercellular tether formation.} Complementary to Results, Fig.5 containing predictions for the remaining models. Simulations with two cells contracting at $50\%$. Cell centers are separated by $6r_{c}$, where $r_{c}$ is the undeformed cell radius.
    \textbf{(a-b)} densification ratio of triangular elements (color plot) in deformed networks \textbf{(d-f)} tensile stretches and \textbf{(g-k)} compressive stretches of deformed fibers.
    \textbf{(m)} Orientation distribution of fibers under tension (stretch $\lambda >1$) within the intercellular region across all models. Each violin corresponds to each one of the models studied and shows the distribution of fiber horizontal direction (in degrees), 
  $^{\ast \ast \ast}p-value< 0.001$\\Colorbars: (a-c) densification ratio of the deformed networks, (d-k) fiber stretch.}
\end{figure}

\begin{figure}[H]
    \includegraphics[width=16cm]{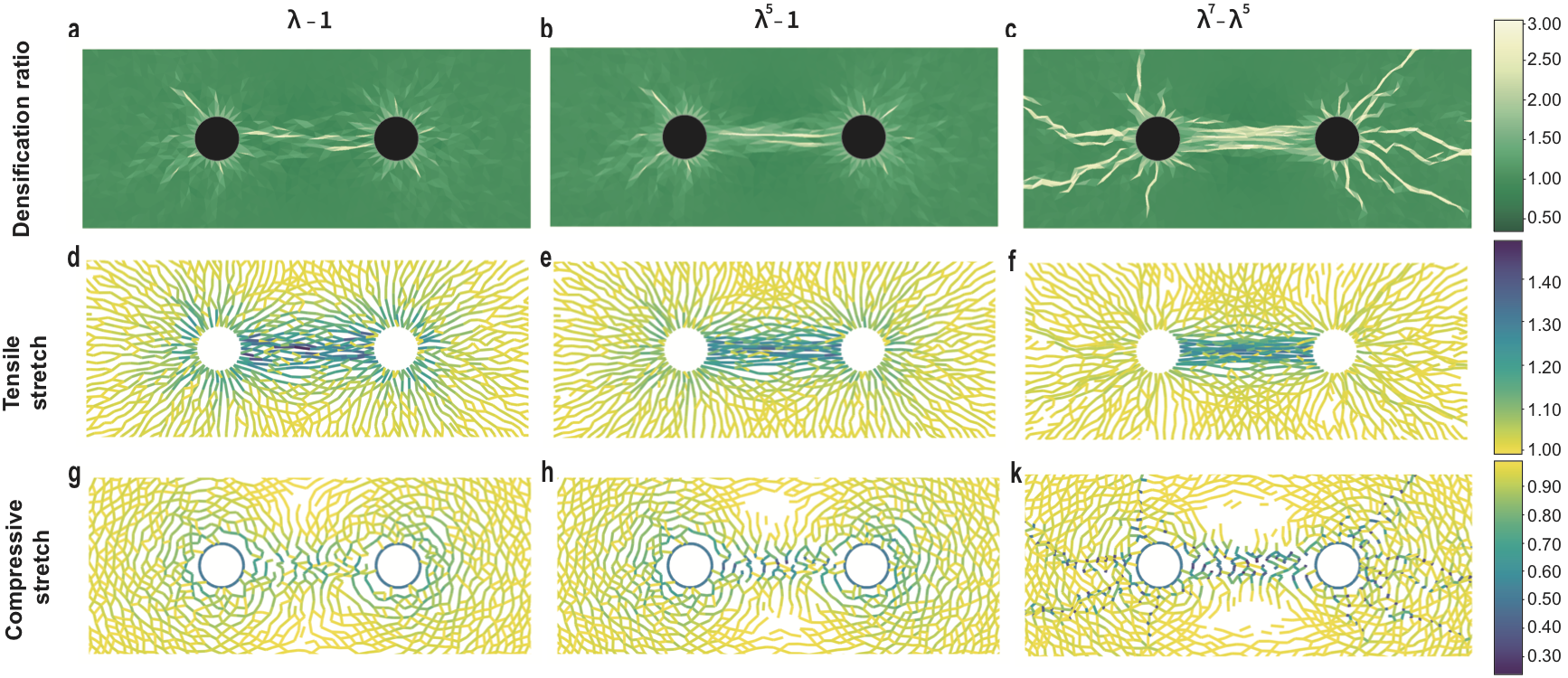}
    \label{sfig6}
    \caption{\textbf{Intercellular tether formation.} Simulations with two cells contracting at $50\%$. As in SFig.5 except that cell centers are separated by $4r_{c}$, where $r_{c}$ is the undeformed cell radius. \textbf{(a-b)} densification ratio of triangular elements (color plot) in deformed networks \textbf{(d-f)} tensile stretches and \textbf{(g-k)} compressive stretches of deformed fibers. We observe densification around each cell boundary, which extends towards the neighbouring cell. Tethers are rather weak for Family-1 cases (a-b) and significantly stronger with Family-2 (c). Within tethers, densification ratio is three times larger than the rest of the matrix. In the intercellular region, fibers under tension are directed towards the neighbouring cell so that they form continual paths connecting the two cells. In these paths, fibers under tension are almost perfectly aligned with the horizontal line connecting the two cells. In Family-2 case (f) excessive tensile stretches are concentrated only within the tether-region. Severely compressed fibers (g-k) locate in the intercellular domain, being roughly perpendicular to fibers under tension. \\Colorbars: (a-c) densification ratio of the deformed networks, (d-k) fiber stretch.}
\end{figure}

\begin{figure}[H]
    \includegraphics[width=17cm]{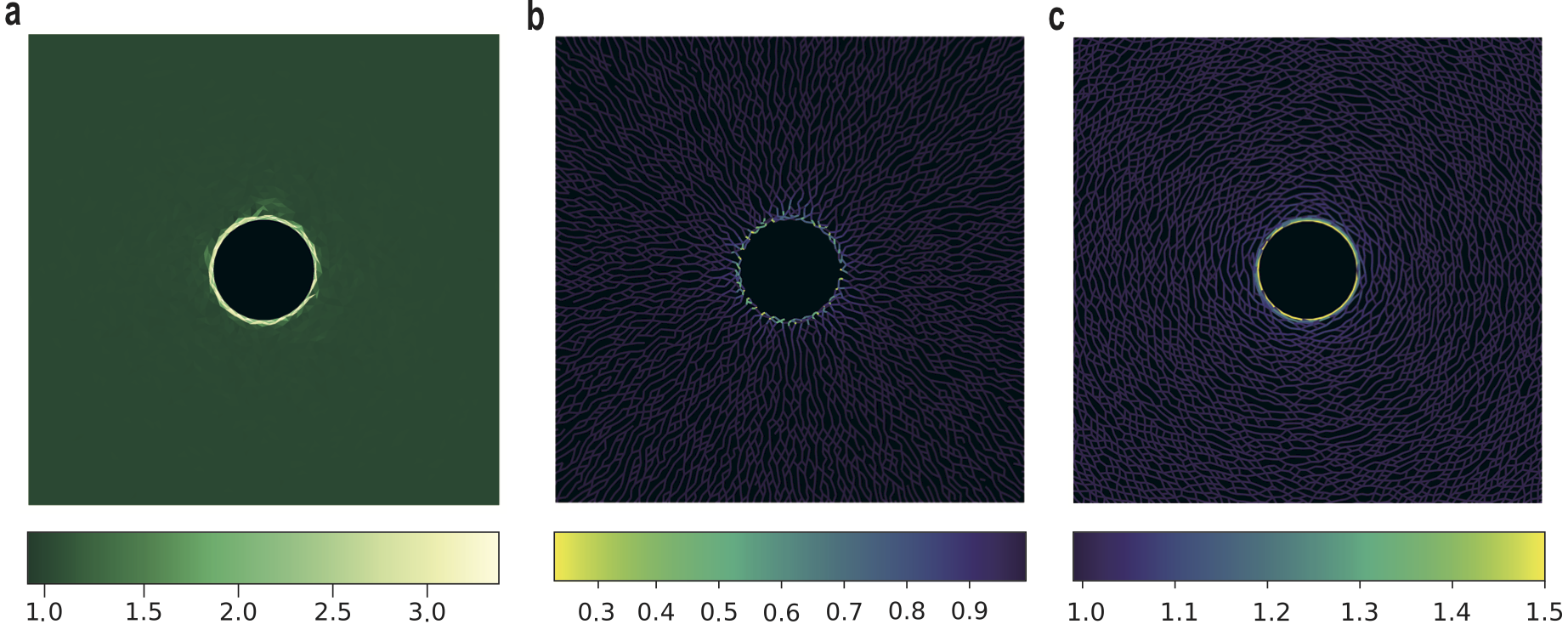}
    \label{sfig7}
    \caption{\textbf{Cell expansion.} Simulation with $S(\lambda) = \lambda^{5} - \lambda^{3}$ of a single cell radially expanded by $50\%$.\textbf{(a)} Densification ratio of triangular elements (color plot) in deformed networks \textbf{(b)} compressive stretches and \textbf{(c)} tensile stretches in deformed fibers. Note that the compressed fibers align with the radial direction while fibers under tension orient in the angular direction.}
\end{figure}


\end{document}